\documentclass[aps,prl,twocolumn,groupedaddress,showpacs,floatfix,superscriptaddress,longbibliography]{revtex4-2}
\usepackage[plainpages=false,pdfpagelabels,colorlinks=true,linkcolor=red,urlcolor=blue,citecolor=blue,pdftitle={Title},pdfauthor={},pdfdisplaydoctitle=true,pdfduplex=DuplexFlipLongEdge]{hyperref}
\usepackage{siunitx}
\usepackage{mhchem}
\usepackage{epsfig}
\usepackage{graphicx}
\usepackage[utf8]{inputenc}
\usepackage{amsmath,amssymb}
\usepackage[dvipsnames]{xcolor}

\usepackage[normalem]{ulem}

\begin{document}

\title{In-depth Investigation of Conduction Mechanism on Defect-induced Proton-conducting Electrolytes BaHfO$_3$ }
\author{Peng Feng}
\affiliation{Frontier Science Center for Flexible Electronics, Xi'an Institute of Flexible Electronics, Northwestern Polytechnical University, 127 Youyi West Road, Xi'an, Shaanxi, 710072, China\\}
\author{Hang Ma}
\affiliation{School of Physics and Astronomy, Beijing Normal University, Beijing 100875, China\\}
\author{Kuan Yang}
\affiliation{Frontier Science Center for Flexible Electronics, Xi'an Institute of Flexible Electronics, Northwestern Polytechnical University, 127 Youyi West Road, Xi'an, Shaanxi, 710072, China\\}
\author{Yingjie Lv}
\affiliation{School of Physics and Astronomy, Beijing Normal University, Beijing 100875, China\\}
\author{Ying Liang}
\email{liang@bnu.edu.cn}
\affiliation{School of Physics and Astronomy, Beijing Normal University, Beijing 100875, China\\}
\affiliation{Key Laboratory of Multiscale Spin Physics (Ministry of Education), Beijing Normal University, Beijing 100875, China\\}
\author{Tianxing Ma}
\affiliation{School of Physics and Astronomy, Beijing Normal University, Beijing 100875, China\\}
\affiliation{Key Laboratory of Multiscale Spin Physics (Ministry of Education), Beijing Normal University, Beijing 100875, China\\}
\author{Jiajun Linghu}
\email{linghujiajun@chd.edu.cn}
\affiliation{Department of Applied Physics, Chang'an University, Xi'an, 710064, China\\}
\author{Zhi-Peng Li}
\email{iamzpli@nwpu.edu.cn}
\affiliation{Frontier Science Center for Flexible Electronics, Xi'an Institute of Flexible Electronics, Northwestern Polytechnical University, 127 Youyi West Road, Xi'an, Shaanxi, 710072, China\\}

\begin{abstract}
This study utilizes first-principles computational methods to comprehensively analyze the impact of A-site doping on the proton conduction properties of BaHfO$_3$. The goal is to offer theoretical support for the advancement of electrolyte materials for solid oxide fuel cells. Our research has uncovered that BaHfO$_3$ demonstrates promising potential for proton conduction, with a low proton migration barrier of $0.28$ eV, suggesting efficient proton conduction can be achieved at lower temperatures. Through A-site doping, particularly with low-valence-state ions and the introduction of Ba vacancies, we can effectively decrease the formation energy of oxygen vacancies (\( E_{\text{vac}} \)), leading to an increase in proton concentration. Additionally, our study reveals that the primary mechanism for proton migration in BaHfO$_3$ is the Grotthuss mechanism rather than the vehicle mechanism. Examination of the changes in lattice parameters during proton migration indicates that while doping or vacancy control strategies do not alter the mode of H$^+$ migration, they do influence the migration pathway and barrier. These findings provide valuable insights into optimizing the proton conduction properties of BaHfO$_3$ through A-site doping and lay a solid theoretical foundation for the development of novel, highly efficient solid oxide fuel cell electrolyte materials.
\end{abstract}

\date{Version 16.0 -- \today}

\maketitle

\section{Introduction}

Clean energy technologies are receiving increasing attention as potential solutions to the pressing challenges posed by climate change and environmental issues. Solid oxide fuel cells (SOFCs) are at the forefront of clean energy conversion technologies, providing a direct and efficient method of converting fuels to electricity with power generation efficiencies in excess of 50\% and combined energy utilization in excess of 80\% with minimal pollution \cite{xu2022comprehensive,bicer2020life}. At the heart of determining SOFC performance is the electrolyte layer between the cathode and anode. Traditional SOFC electrolytes are oxygen-ion conductors, such as yttrium-doped zirconia stabilized by yttrium oxide (YSZ), which operates optimally at temperatures above 800~\si{\celsius} \cite{timurkutluk2016review,ryu2023nanocrystal}. In recent years, proton-conducting SOFCs (H-SOFCs) have become the most attractive option for large-scale energy storage and conversion, and proton-conducting oxides show particular potential due to their small activation energies at intermediate temperatures and ionic conductivity comparable to or even higher than that of oxy-ionic conductors, which allows them to operate efficiently at low to intermediate temperatures (400--600~\si{\celsius}) \cite{ding2020self}, facilitating an efficient, effective and efficient transition between clean, efficient, economical, and long-lasting conversion between electricity and green hydrogen \cite{saito2023high}.

A variety of materials structured with proton-conducting properties have been reported, such as H-doped SrCoO$_{2.5}$ with a proton conductivity close to 0.33~\si{S/cm} (140~\si{\celsius}) and SmNiO$_3$ also with a proton conductivity of 0.03~\si{S/cm} (140~\si{\celsius}) \cite{lu2022enhanced,islam2020computational}. However, the best performance of these types of proton conductors is in the low temperature domain, leading to their composition of batteries that require additional noble metal catalysts to drive the fuel catalysis. At the same time, the degenerate nature of the B-site transition metals tends to make the materials simultaneously highly electronically conductive, leading to a reduction in the open-circuit voltage of the cell \cite{ozawa2018intercalated}. Therefore, Ba-based chalcogenide oxides have become a hot spot in the research of H-SOFC electrolyte materials, among which a series of electrolyte materials, such as BaZrO$_3$ \cite{sun2014easily}, BaCeO$_3$ \cite{gong2018barium}, and their respective solid solutions \cite{lyagaeva2016new}, have been widely investigated and achieved better results. However, BaCeO$_3$ and BaZrO$_3$ also have obvious drawbacks. BaCeO$_3$ is chemically unstable in the presence of H$_2$O and CO$_2$, which is a major problem for practical applications that require long-term operation \cite{kannan2013chemically}. BaZrO$_3$ is much more chemically stable but has a slow growth rate of the grains, which leads to a high concentration at the grain boundaries, thus impairing the overall electrical conductivity \cite{han2018best}. Up to now, the proton conductivity of Ba-based chalcogenide electrolyte materials is still on the order of 10$^{-2}$--10$^{-4}$~\si{S/cm}, which is a large gap with the performance of oxygen-ion-conducting electrolytes ($\geq$10$^{-2}$~\si{S/cm}), and its limited proton transport performance has hindered the widespread application and industrial-scale production of H-SOFC \cite{hossain2017review}. BaHfO$_3$ is a Ba-based chalcogenide oxide isostructured with BaZrO$_3$, in which Hf is in the same group as Zr \cite{bandura2010hybrid}. Kang and Sholl \cite{kang2017characterizing} and Luo \cite{luo2024harnessing} et al. have demonstrated the potential of BaHfO$_3$ as a proton-conducting electrolyte material in high-throughput calculations on Ba-based chalcogenides. Also, the better corrosion resistance of BaHfO$_3$ provides the basis for its stability. In addition, the proton migration energy barrier of BaHfO$_3$ is predicted to be about 0.28~\si{eV}, which is lower than that of an electrolyte such as BaZr$_{0.875}$Y$_{0.125}$O$_3$ at 0.42~\si{eV} \cite{gomez2012periodic}, indicating its high proton transport performance.

The proton transport process is closely related to the proton concentration and transport rate in the electrolyte, and the regulation of its performance needs to take both into account. Ba-based chalcogenides are often doped with B-site doping to introduce oxygen vacancies, which enhances the proton entry into the electrolyte water and the reaction process, thus increasing the proton concentration into the electrolyte, but also due to the doping ions' trapping effect on the protons and blocking the proton transport rate \cite{bu2016effect,guo2022sn,oh2024novel,wang2024large}. In contrast, A-site defect induction can also introduce oxygen vacancies in the material, while the trapping effect on protons is weaker than that of B-site doping, which allows for finer regulation of the proton transport process \cite{ivanova2012effects}.

There is a lack of systematic studies on Ba-based chalcogenides, especially on the A-site defect induction of BaHfO$_3$ with respect to the proton concentration and transport rate, so the present study analyzes the performance of BaHfO$_3$ as an H-SOFC electrolyte by providing a systematic first-principles analysis of the A-site defect induction of BaHfO$_3$ \cite{li2011structural}. The study investigated the oxygen vacancy formation energy and proton transport barrier of BaHfO$_3$ in different A-site states (isovalent and low-valent doping, and the A-site vacancy scenarios were set up to take into account the phenomenon that Ba$^{2+}$ is prone to volatilization during the actual preparation process), and compared the intrinsic connection between the main influencing factors affecting the proton concentration and transport rate, i.e., the oxygen vacancy concentration and the proton transport energy barrier, and the different states of the A site in the different scenarios. Our results confirm the great potential of BaHfO$_3$ as an H-SOFC electrolyte and are also crucial for understanding the proton transport process in the electrolyte and probing the proton transport mechanism.

\section{Model and method}

Density functional theory (DFT) with the projector augmented wave (PAW) \cite{blochl1994projector} method and Perdew-Burke-Ernzerhof (PBE) based generalized gradient approximation (GGA) \cite{perdew1996generalized} functional was performed with the Vienna Ab Initio Simulation Package (VASP) \cite{kresse1996efficient,min2022first}. The cutoff energy of the plane wave basis was set to 520~eV. The Monkhorst-Pack method \cite{monkhorst1976special} was employed for k-point sampling, with a spacing of 0.028~\si{\angstrom^{-1}}. Structural optimizations were carried out using the conjugate gradient algorithm \cite{yu2009first} and were performed with the self-consistency precision energy of $10^{-7}$~eV until the residual force was converged to less than 0.01~\si{eV/\angstrom}. All the simulation calculations were achieved via a $2 \times 2 \times 2$ supercell. 

Proton migration barriers were calculated using the climbing image nudged elastic band (CI-NEB) method \cite{henkelman2000climbing}. A total of nine images were interpolated along the minimum energy path between the initial and final proton positions. Structural optimization of each image was performed with the volume and shape of the supercell kept constant, while allowing only the atomic coordinates to relax. Convergence criteria of $10^{-5}$~eV and $10^{-2}$~\si{eV/\angstrom} were used for energy and force, respectively.

\section{Results and Discussion}

\begin{figure}[htbp]
    \centering
    \includegraphics[scale=0.5]{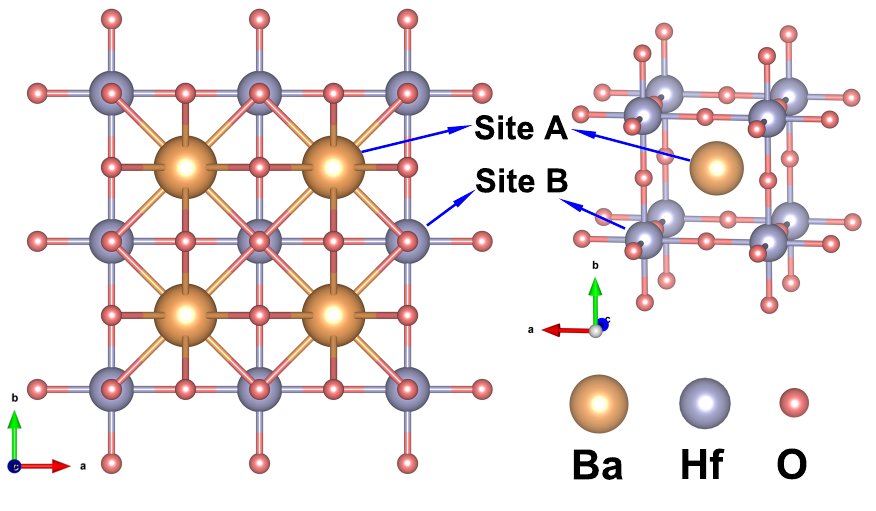}
    \caption{\label{Fig1}Structure of BaHfO$_3$.}
    \end{figure}

As illustrated in FIG.~\ref{Fig1}. BaHfO$_3$ has a cubic chalcogenide structure and belongs to the tetragonal $Pm\mathop{3}\limits^{-}m$space group\cite{guevara1998structure}. Hf$^{4+}$ ions combine with six O$^{2-}$ atoms to form [HfO$_6$] octahedra. These octahedra are co-angled with six equivalent octahedra, and the Ba ions populate the space enclosed by the [HfO$_6$] octahedra. The lattice constant of BaHfO$_3$ calculated in this study is $a = b = c = 4.20$~\si{\angstrom}, which is very close to the value in the Database Materials Project ($a = b = c = 4.17$~\si{\angstrom}). Meanwhile, the six Hf-O bond lengths in the [HfO$_6$] octahedron are all approximately 2.08~\AA, with O-Hf-O bond angles close to $90^{\circ}$. Therefore, it can be inferred that the symmetry of the [HfO$_6$] octahedra in BaHfO$_3$ is exceptionally high, so that each octahedron will provide the same environment for proton migration, facilitating efficient proton transport.
  
As mentioned before, the proton conductivity of the electrolyte material in SOFC depends on two factors: the proton concentration in the electrolyte and the proton transport velocity in the electrolyte.  In Ba-based electrolyte materials, the proton concentration depends strongly on the following hydration process that \cite{lee2022defect}
\begin{equation}
\ce{H2O + O_{o} + V^{**}_{o} -> 2OH^{*}_{o}}
\label{Eq1}
\end{equation}

Where O$_o$ and V$^{**}_{o}$ in Eq. (\ref{Eq1}) represent lattice oxygen and oxygen vacancy, respectively, and OH$^{*}_{o}$ represents a proton defect formed by the interaction of a proton and a lattice oxygen.

\begin{figure}[htbp]
\centering
\includegraphics[scale=0.5]{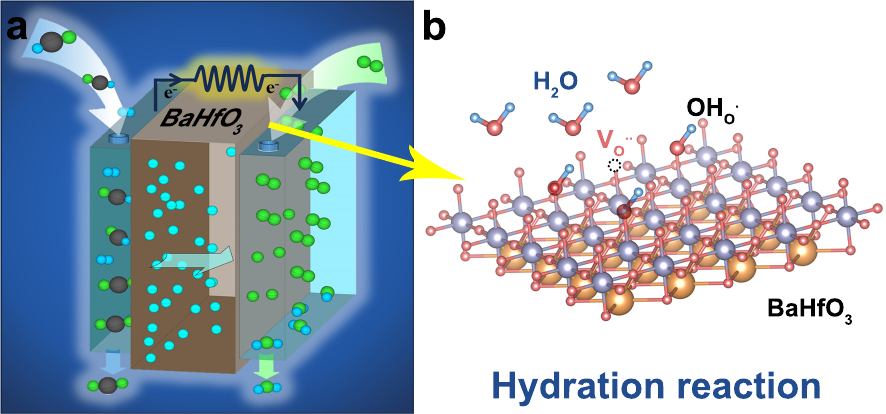}
\caption{\label{Fig2}(a) Structure of SOFC and power generation process, and (b) electrolyte attracts protons through water and reaction.}
\end{figure}

In turn, the efficiency of the hydration process is controlled by the oxygen vacancies in the electrolyte (e.g., FIG.~\ref{Fig2}). Therefore, we first evaluated the formation energy of various defects ($E_\text{defect}$, FIG.~\ref{FigS1}) and the generation of nearest neighbor (T-N) and next-nearest neighbor (T-NN) oxygen vacancies (FIG.~\ref{FigS2}a) in Ba$_{0.875}$M$_{0.125}$HfO$_3$ under different scenarios (M=substrate, A-site substitution or vacancy). The situation will be judged on the basis of the oxygen vacancy formation energy, which is calculated by using the following equation \cite{ganduglia2004stability}.

\begin{equation}
E_{vac}=E_{withV_{o}}+\frac{1}{2}E_{O_2}-E_{withoutV_{o}}
\end{equation}
where \( E_{\text{vac}} \) denotes the formation energy of oxygen vacancies, \( E_{\text{withV$_{o}$}} \) denotes the energy of a system (perfect BaHfO$_3$, or with A-site defect) containing an oxygen vacancy, and \( E_{\text{O$_2$}} \) is the chemical potential of oxygen.  The energy is expressed using \(E_{\text{withoutV$_{o}$}}\) for a pristine or defective crystal without \(V_{o}\).

\begin{figure}[htbp]
\centering
\includegraphics[scale=0.53]{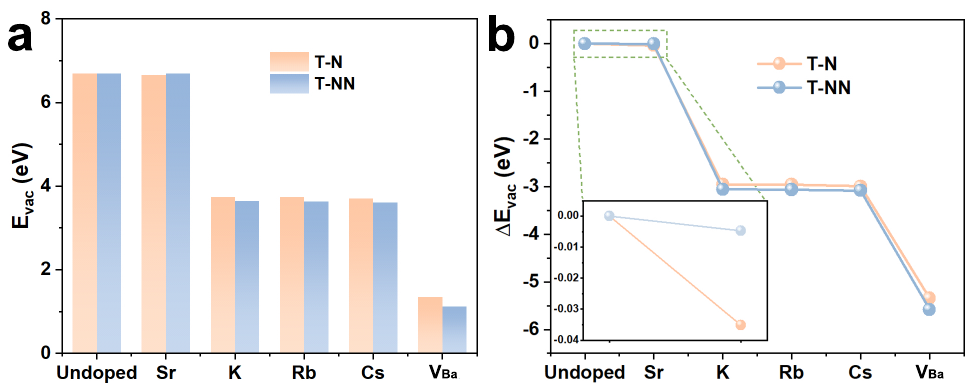}
\caption{\label{Fig3}(a) $E_{vac}$ and (b) $\Delta E_\text{vac}$ of T-N and T-NN under different A-site modulations.}
\end{figure}

The results are shown in FIG.~\ref{Fig3}(a), where the formation energies of oxygen vacancies in perfect BaHfO$_3$ are 6.69~eV. When 12.5\% of Ba$^{2+}$ are replaced by the homovalent Sr$^{2+}$, the $E_\text{vac}$ slightly decreases to about 6.66~eV and 6.68~eVfor the vacancy located on the nearest (T-N) and next nearest (T-NN) position to the dopant, respectively. In contrast, after the monovalent ion substitution of Ba$^{2+}$, the $E_\text{vac}$ of Ba$_{0.875}$M$_{0.125}$HfO$_3$ (M=K$^{+}$, Rb$^{+}$, Cs$^{+}$) decreases below 4~eV, and the $E_\text{vac}$ of T-N and T-NN after the introduction of the Ba vacancy is 1.35~eV and 1.12~eV, respectively. The decrease of the $E_\text{vac}$ ($\Delta E_\text{vac}$) under the Sr$^{2+}$ doping of T-N and T-NN  is only about 0.04~eV and 0.01~eV, and after substitution with monovalent ions, $\Delta E_\text{vac}$ is close to or exceeds 3 eV. $\Delta E_\text{vac}$ even exceeds 5.3 eV when Ba vacancy is introduced(FIG.~\ref{Fig3}(b)).

According to the defect equation, the generation of oxygen vacancies in the case of eigenstates, doping, and Ba vacancies is given by
\begin{equation}
\ce{BaHfO_3 -> V_o^{**} + \frac{1}{2}V_{Ba}^{''} + \frac{1}{4}V_{Hf}^{''''}} 
\label{Eq3}
\end{equation}

\begin{equation}
\ce{Ba_{Ba} + A^{+} -> A_{Ba}^{'} + \frac{1}{2}V_{o}^{**}} 
\label{Eq4}
\end{equation}

\begin{equation}
\ce{Ba_{Ba} + Sr^{2+} -> Sr_{Ba} + Ba^{2+}} 
\label{Eq5}
\end{equation}
where \( V_{O}^{**} \) denotes oxygen vacancies, \( V_{Ba}^{''} \) denotes Ba vacancies, and \( A \) denotes monovalent ions (\( A^+ = \text{K}^+, \text{Rb}^+, \text{Cs}^+ \)). The defect equation discussed above clearly shows that in Eq. (\ref{Eq3}), the introduction of Ba vacancies causes the crystal to prompt the production of an equal amount of \( V_{O}^{**} \) (FIG.~\ref{FigS2}(c)). In the case of Eq. (\ref{Eq4}), the crystal would be in a state of charge imbalance due to the substitution of low-valence ions, thus driving the charge compensation effect and the generation of \( V_{O}^{**} \). This is consistent with the apparent decrease in \( E_{\text{vac}} \) observed after substitution of Ba vacancies and monovalent ions in FIG.~\ref{Fig3}. In the case of substitution of Ba\(^{2+}\) by Sr\(^{2+}\) in the homovalent state, i.e., the case described in Eq. (\ref{Eq5}), there is no driving force for oxygen vacancy formation originating from the charge effect, and thus theoretically no oxygen vacancies are generated.However, according to the conclusion of FIG.~\ref{Fig3}, \( E_{\text{vac}} \) also exhibits a small decrease after Sr\(^{2+}\) substitution, which is due to the fact that Sr\(^{2+}\) and Ba\(^{2+}\) have different ionic radii although they have the same valence state. The ionic radius at 12 coordination numbers (\( N \)) of Sr\(^{2+}\) (1.44~\AA) is smaller than that of Ba\(^{2+}\) (1.61~\AA). The smaller Sr\(^{2+}\) ions could not completely fill the space of Ba\(^{2+}\), making the lattice unstable for oxygen, and the lattice distortion due to this radius difference promoted the formation of \( V_{O}^{**} \) (FIG.~\ref{FigS2}(b)). However, in the doping of monovalent ions, K\(^{+}\) ( $R_{\text{K}^+}$ = 1.64~\AA, N=12 ), which has a similar radius to Ba\(^{2+}\), also leads to a substantial \( E_{\text{vac}} \) decrease. The \( E_{\text{vac}} \) gradually decreased with the increase of doped ion radius ( $R_{\text{Rb}^+}$ = 1.72~\AA, $R_{\text{Cs}^+}$ = 1.88~\AA, N=12 ) (FIG.~\ref{FigS2}(d)). The above results suggest that the charge compensation effect due to ion valence is the main driving force for oxygen vacancy formation when the ion radii are similar, while the difference in ion radii (either on the larger or smaller side) at the same valence state contributes to the generation of oxygen vacancies.

The analysis \( E_{\text{vac}} \) shows that a large number of oxygen vacancies are formed in the material after the doping of low valence ions and the introduction of Ba vacancies, and therefore a high concentration of protons will exist in both cases.

\begin{figure*}[htbp]
    \centering
    \includegraphics[scale=0.5]{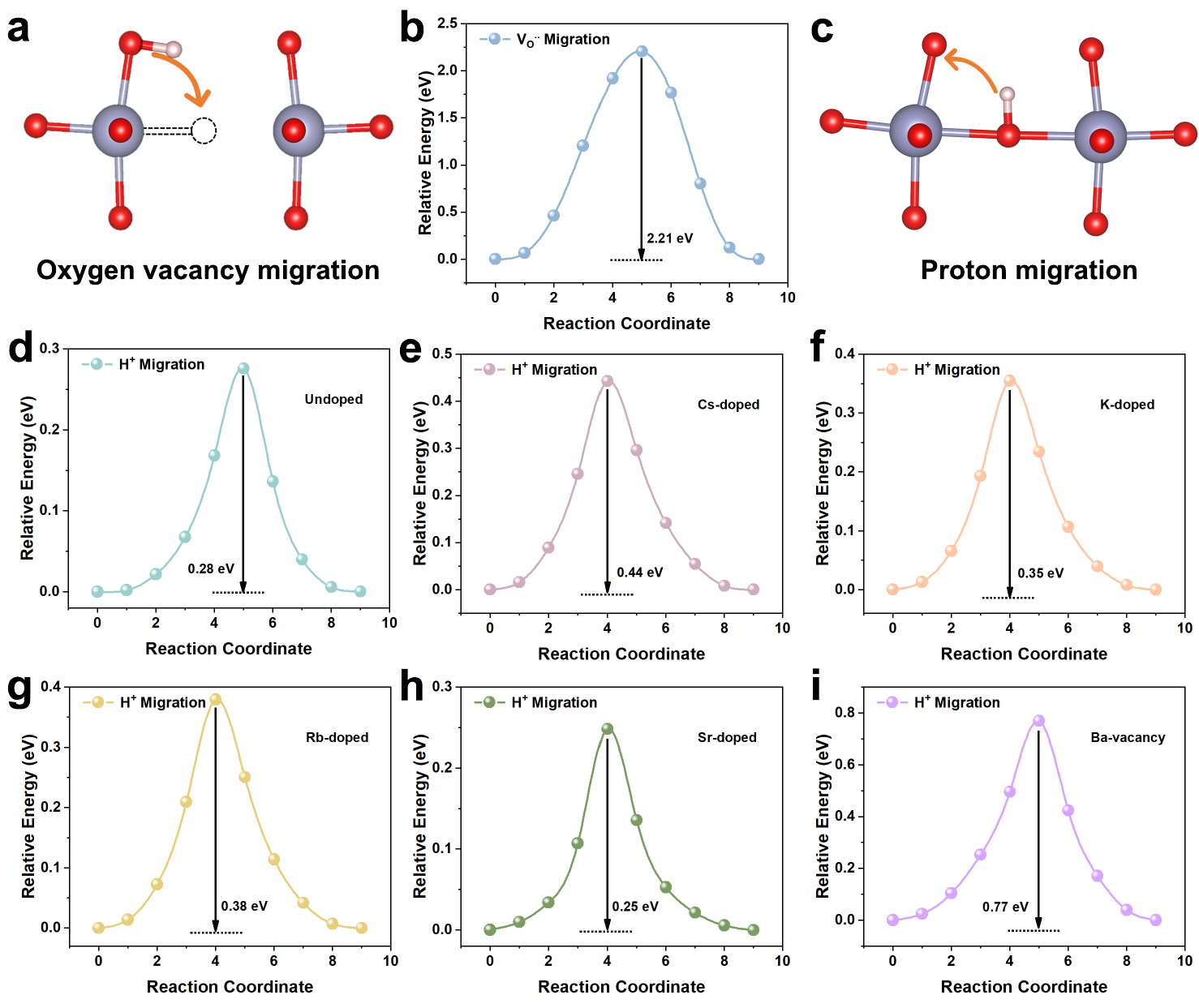}
    \caption{\label{Fig4}(a) Vehicle mechanism in the  migration process with oxygen vacancies. (b) Migration energy barrier of oxygen vacancies. (c) Proton migration process. (d) Undoped, (e) Sr doping, (f) K doping, (g) Rb doping, (h) Cs doping, and (i) proton migration energy barriers at Ba vacancies.}
    \end{figure*}

As mentioned earlier, another factor affecting the proton conduction performance is the proton transport velocity, which is directly related to the existence of paths in the lattice that satisfy the high proton migration speed. Therefore, we simulated the proton migration paths and the corresponding energy barriers for all the above cases.Considering that there are two hypothetical models of proton conduction: One is the vehicle mechanism in which H$^+$ forms proton defects with O$^{2-}$ in the lattice and migrates in the lattice with the exchange of oxygen ions and oxygen vacancies (FIG.~\ref{Fig4}(a)), and the other is the Grotthuss mechanism in which protons migrate in the form of a spinning-jumping pattern with the oxygen of the lattice as the backbone (FIG.~\ref{Fig4}(c)). The main migration mechanism in the material is determined by the simulation of the migration barriers of oxygen vacancies and protons on the set path.The calculated migration energy barrier for oxygen vacancies in FIG.~\ref{Fig4}(b) is about 2.21~eV, indicating that the energy to be overcome for the exchange migration of oxygen ions of larger radius with oxygen vacancies is high, and thus the vehicle transport mechanism of protons is not dominant in BaHfO$_3$. In contrast, calculations based on the Grotthuss mechanism show that the migration energy barrier of protons is much lower than that of oxygen ions. FIG.~\ref{Fig4}(d) exhibits a proton migration barrier of 0.28~eV in the undoped perfect BaHfO$_3$ lattice, and the equivalent Sr$^{2+}$ doping further reduces the proton migration barrier to 0.25~eV (FIG.~\ref{Fig4}(e)). For monovalent doping, the proton mobility barrier gradually increases from 0.35~eV to 0.44~eV with the increase of the radius of the low-valent dopant ions (K$^+$ $\rightarrow$ Cs$^+$) as shown in FIG.~\ref{Fig4}(f)-(h). The mobility barrier of H$^+$ in the presence of Ba vacancies is 0.77~eV (FIG.~\ref{Fig4}(i)).

Combined with the previous calculations, we found that the trends of \( E_{\text{vac}} \) and proton mobility barriers in the above A-site defect induction process are completely different. \( E_{\text{vac}} \) has been showing a decreasing trend in the case of A-site defect induction and vacancies, while the proton mobility barriers are higher than those of BaHfO$_3$ except for a slight decrease in the case of Sr$^{2+}$ doping. Therefore, after comparing the magnitude of the change in \( E_{\text{vac}} \) and proton migration energy barrier in the A-site defect-induced mode, it was found that the doping of  homovalent Sr$^{2+}$ was mainly manifested in the effect on the proton migration energy barriers, whereas the doping of low valence ions had a more significant effect on the change in \( E_{\text{vac}} \) (FIG.~\ref{FigS3}). This indicates that the doping of low valence state ions is an effective measure to improve the proton conduction properties of BaHfO$_3$ by substantially increasing the proton concentration in BaHfO$_3$.

\begin{figure*}[htbp]
    \centering
    \includegraphics[scale=0.6]{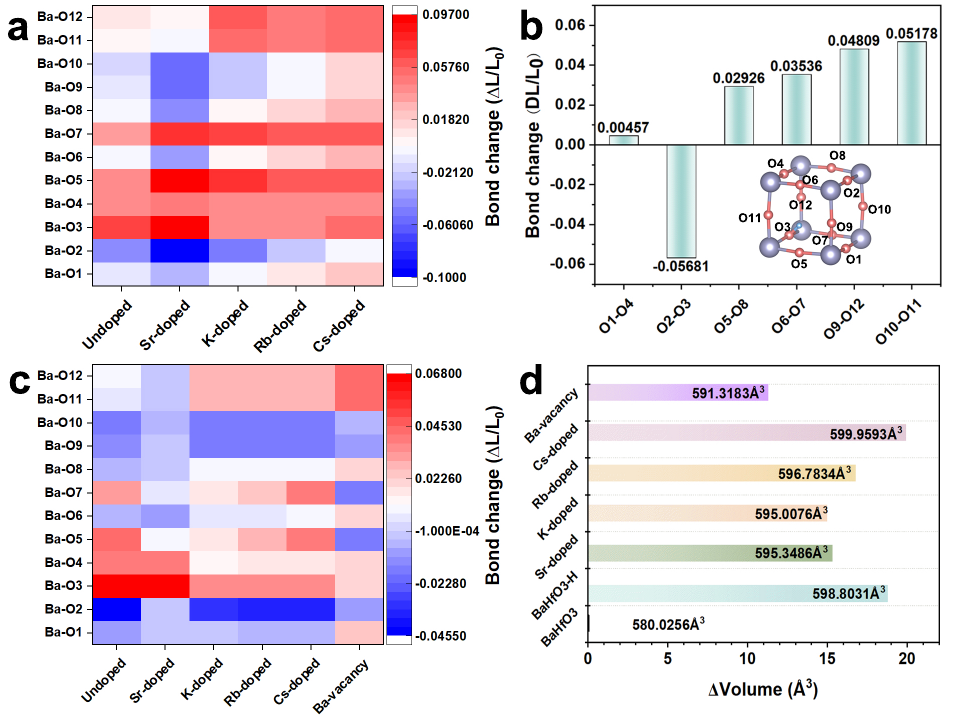}
    \caption{\label{Fig5}a. The bond length changes between the substituent and the surrounding coordinating oxygen, b. The bond length between oxygen atoms is altered in the presence of a vacancy in the Ba site. c. The bond length changes between the adjacent Ba site and the coordinated oxygen. d. Changes in the volume of the supercell under different A-site defect induction.}
    \end{figure*}

Differences in ionic radii can lead to lattice distortions, which are specifically manifested as variations in bond lengths and unit cell volumes. By examining the simulation results of the unit cells during the A-site defect induction process for various samples (FIG.~\ref{FigS4}), we observed changes in the bond lengths and unit cell volumes between the substitution sites and adjacent sites with their coordinated oxygens, as depicted in FIG.~\ref{Fig5}. The degree of bond length change was assessed using the following formula,

\begin{equation}
\Delta Bond = \frac{L_j-L_0}{L_0} 
\end{equation}

From FIG.~\ref{Fig5}(a) and \ref{Fig5}(b), it is evident that, compared to the perfect BaHfO$_3$ unit cell with H$^+$ transfer in it, all A-site defect-induced [MO$_{12}$] polyhedra exhibit expansion except for the [SrO$_{12}$] polyhedron after Sr doping, which shows contraction (with an average bond length of 2.94336 \AA). This includes the [V$_{Ba}$O$_{12}$] polyhedron in the case of Ba vacancy (where there is no central atom, the distortion of the polyhedron is assessed by the changes in bond lengths between the apical oxygens), as demonstrated by the data in Tables \ref{Table S1} and \ref{Table S2}. Among them, the ionic radius increases from K$^+$ to Cs$^+$, and the degree of expansion of the [MO$_{12}$] also increases. Excluding the special case of Ba vacancy, an analysis was conducted to correlate the distortion of the [MO$_{12}$] polyhedral, represented by the average bond length variations, with the ionic radii and proton migration barriers. It was observed that these three parameters exhibit a linear relationship with one another, as illustrated in FIG.~\ref{FigS5}. Comparatively, FIG.~\ref{Fig5}(c) demonstrates that while there are minor fluctuations in the bond lengths between adjacent Ba sites and their coordinated oxygens, the average bond length alterations are negligible ($\leq0.01$ \AA, as depicted in Table \ref{Table S3}). These findings suggest that the degree of distortion of the polyhedral at the substitution sites post-ionic doping has a direct impact on the proton migration barriers.

As shown in FIG.~\ref{Fig5}(d), the introduction of protons causes the volume of the BaHfO$_3$ matrix crystal to expand, while the subsequent continued insertion of equivalent Sr$^{2+}$, K$^{+}$, and Rb$^{+}$, which have smaller, similar, and larger ion radii, respectively, causes the unit cell volume to decrease. This reduction in the case of Sr$^{2+}$ is due to lattice contraction stemming from the decreased ionic radius, which consequently shortens the H$^+$ migration distance and diminishes the migration energy. For K$^+$ and Rb$^+$, the charge-induced shift of H$^+$ toward the A site, despite a reduction in volume, leads to an increase in migration energy due to the attractive force exerted by the A-site ions on H$^+$. With Cs$^{+}$ doping, the larger ionic radius of Cs$^{+}$ (1.88~\AA) leads to a lattice expansion and stronger electrostatic repulsion, which further raises the migration energy. In the scenario of Ba vacancy, although there is a substantial contraction in unit cell volume, the absence of A-site ions intensifies the binding effect on H$^+$, thereby complicating proton migration. These findings elucidate that lattice distortion and charge effects are the predominant factors contributing to changes in migration energy. Lattice distortion influences the migration energy barrier by affecting the proton migration distance, whereas the charge effect determines the driving force for proton migration based on the magnitude of its binding to the proton.

\begin{figure*}[htbp]
    \centering
    \includegraphics[scale=0.5]{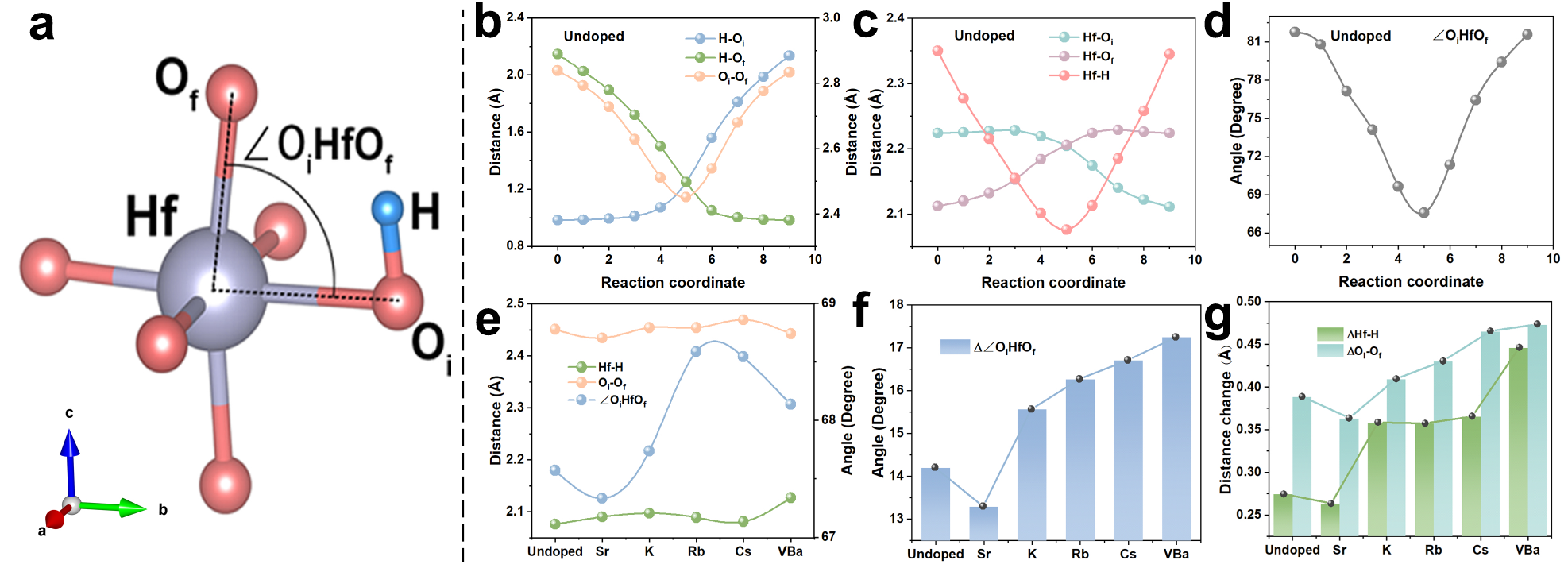}
    \caption{\label{Fig6}(a) Schematic diagram of the HfO$_6$ octahedron associated with the H undoped BaHfO$_3$ during proton migration. (b) H-O$_i$, H-O$_f$, and O$_i$-O$_f$. (c) Distances of Hf-O$_i$, Hf-O$_f$, and Hf-H. (d)  Angular variations of $\angle$O$_i$HfO$_f$. (e) Variations of Hf-H, O$_i$-O$_f$, and $\angle$O$_i$HfO$_f$ at different A-site defect induction modes at the initial state. (f),(g) Variations of Hf-H, O$_i$-O$_f$, and $\angle$ O$_i$HfO$_f$ at different A-site defect induction modes at the initial and critical states.}
    \end{figure*}

In order to probe deeply into the reasons leading to the above results, the changes of lattice parameters, including bond lengths and bond angles, during proton migration are analyzed in detail. As an example, an [HfO$_6$] octahedron during proton migration is shown in FIG.~\ref{Fig6}(a). where H$^+$ migrates from the initial oxygen O$_i$ along the $b \times c$ plane of the lattice, to the final oxygen O$_f$ as the set migration path. The bond lengths associated with this process, such as H-O$_i$, H-O$_f$, O$_i$-O$_f$, H$_f$-O$_i$, H$_f$-O$_f$, and H$_f$-H, as well as the bond angle $\angle$O$_i$HfO$_f$, all change synchronously. Figure~\ref{Fig6}(b) demonstrates that in undoped BaHfO$_3$, $\angle$O$_i$HfO$_f$ decreases with the shortening of the Oi-Of bond and the gradual rotation of H$^+$ toward O$_f$. The H-O$_i$ bond gradually elongates during this process and reaches the equality of the H-O$_i$ and H-O$_f$ bond lengths (1.25 \AA) at the shortest point of the O$_i$-O$_f$ (2.45 \AA), when the $\angle$O$_i$HfO$_f$ also reaches the minimum value of $67.57^{\circ}$. At this critical point, the H-O$_i$ bond breaks, and H$^+$ breaks away from O$_i$ and bonds with O$_f$. Thereafter, the H-O$_i$ distance increases and the H-O$_f$ bond is gradually shortened, thus realizing a single migration of H$^+$ from O$_i$ to O$_f$ (FIG.~\ref{Fig6}(b) and \ref{Fig6}(d)). During the migration process, the Hf-O$_i$ bond shortens, while the Hf-O$_f$ bond gradually elongates in order to maintain the stability of the octahedral energy; the two bond lengths are equal at the critical point (2.20 \AA), and the distance between Hf and H reaches a minimum of 2.08 \AA. The Hf-O$_i$ bond shortens to 2.08 \AA, and the Hf-O$_f$ bond gradually elongates to 2.20 \AA.

In the context of proton conduction mechanisms, the trends observed with doping and Ba vacancy are largely consistent with those in the undoped material (FIG.~\ref{FigS6}). This suggests that the doping or vacancy modulation strategy does not alter the "rotational" mode of H$^+$ migration but rather affects the length of the proton migration pathway. To substantiate this inference, further comparisons were made regarding the structural deformation under various conditions. The comparative results in FIG.~\ref{Fig6}(e) reveal that the initial O$_i$-O$_f$ length and the angle $\angle$O$_i$HfO$_f$ exhibit a trend of decreasing, then increasing, and then decreasing again. Upon examining the [HfO$_6$] octahedra in all materials, it was observed that in the undoped BaHfO$_3$, both the [HfO$_6$] octahedra and the coordinated oxygen with H$^+$ are located in the $b \times c$ plane. With the doping of smaller radius Sr$^{2+}$, the [HfO$_6$] octahedra tilt toward the Sr$^{2+}$, the $\angle$O$_i$HfO$_f$ decrease, and the O$_i$-O$_f$ distance shortens. When larger radius monovalent ions are doped, the tilt of the [HfO$_6$] octahedra decreases, and even after doping with Cs$^+$, the [HfO$_6$] octahedra tilt in the opposite direction, at which point the O$_i$-O$_f$ bond distance elongates. In the case of Ba vacancy, the surrounding [HfO$_6$] octahedra contract toward the vacancy, thus shortening the O$_i$-O$_f$ bond length and $\angle$O$_i$HfO$_f$, depicted in FIG.~\ref{FigS6}. The Hf-H distance, however, shows an opposite trend to the former two, due to the changes in the length of the Hf-O$_i$ bond and the angle with the H-O$_i$ bond in different materials. When Ba vacancy is introduced, the Hf-H distance exhibits the maximum length (2.22 \AA), at which point H$^+$ deflects significantly toward the Ba vacancy. This is attributed to the charge imbalance at the A site caused by the Ba vacancy, thereby promoting the shift of H$^+$ toward the A site, as illustrated in FIG.~\ref{FigS6}(a)-\ref{FigS6}(f).

FIG.~\ref{Fig6}(f) and FIG.~\ref{Fig6}(g) depict the $\Delta\angle$O$_i$HfO$_f$, $\Delta$Hf-H, and $\Delta$O$_i$-O$_f$ during the initial and transition states of H$^+$ transfer. From these figures, it can be observed that when BaHfO$_3$ is doped with Sr$^{2+}$ of the same valence, the changes in all three parameters are reduced, indicating that the [HfO$_6$] octahedron does not need to undergo significant deformation to complete the migration. This implies that the length of the proton migration pathway is shortened, which is well in line with the observed reduction in proton migration barriers upon Sr$^{2+}$ doping. In contrast, when K$^+$, a monovalent ion with a similar radius, substitutes for Ba$^{2+}$, the changes in the three parameters increase. Specifically, $\Delta\angle$O$_i$HfO$_f$ and $\Delta$O$_i$-O$_f$ increase by $15.6\%$ and $5.4\%$, respectively, compared to undoped BaHfO$_3$, while $\Delta$Hf-H increases by $30.7\%$. This suggests that the substitution of monovalent ions at the A-site significantly affects the relative distance between the initial and transition states of H$^+$, followed by the degree of distortion of the [HfO$_6$] octahedron.

Comparing with the sample substituted by K$^+$, the changes in the three parameters after doping with Rb$^+$ and Cs$^+$ show that $\Delta\angle$O$_i$HfO$_f$ and $\Delta$O$_i$-O$_f$ increase with the ion radius, but $\Delta$Hf-H does not show a significant increase. This implies that doping with monovalent ions of different radii primarily affects the distortion of the [HfO$_6$] octahedron and does not significantly impact the relative distance between the initial and transition states of H$^+$. In the case of Ba vacancy, compared to the undoped situation, the changes in $\Delta\angle$O$_i$HfO$_f$ and $\Delta$O$_i$-O$_f$ are $17.2\%$ and $21.9\%$, respectively, which are much smaller than the $62.8\%$ change in $\Delta$Hf-H. Therefore, the vacancy has a more significant impact on the relative distance between the initial and transition states of H$^+$ (as shown in FIG. ~\ref{FigS7} and \ref{FigS8}). These results fully demonstrate that the proton migration pathway consists of two parts: first, the displacement of the proton as the coordinated oxygen moves, and second, the displacement of the proton itself. Changes in the radius of the A-site ions predominantly govern the first type of displacement, while the second type is more attributable to changes in the ion valence.

\begin{figure*}[htbp]
    \centering
    \includegraphics[scale=0.5]{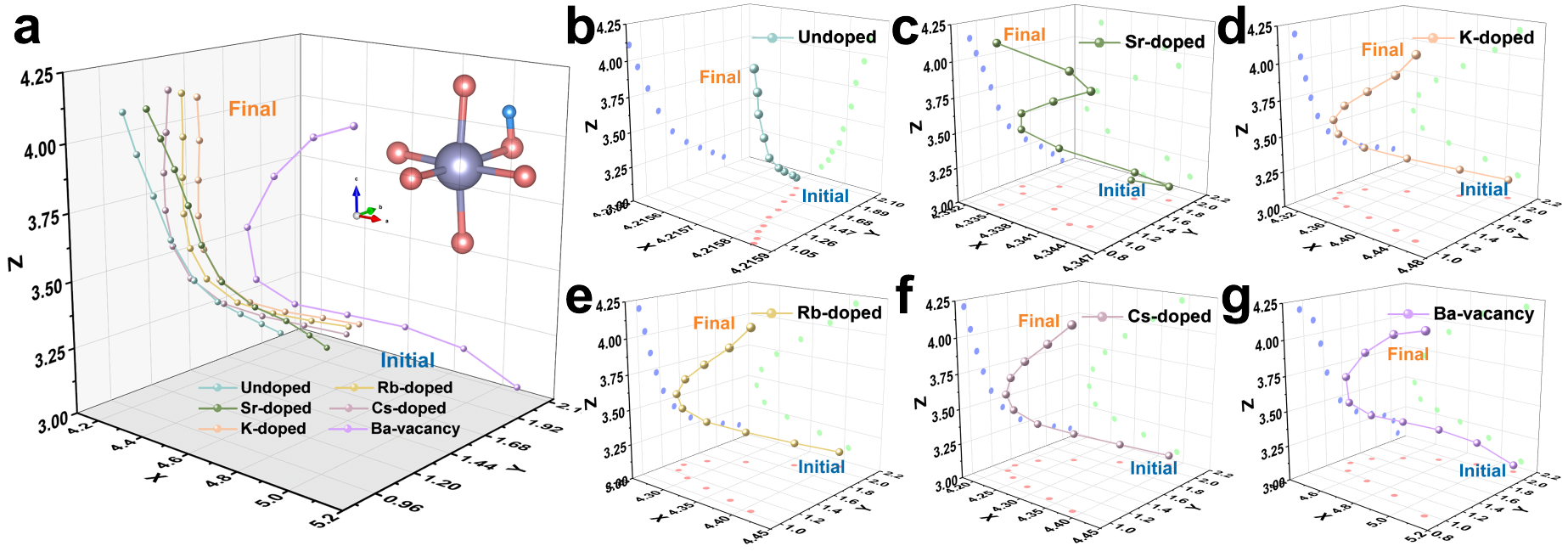}
    \caption{\label{Fig7}(a) Migration paths of protons from O$_i$ to O$_f$ under different A-site regulation methods. The migration paths of protons in the (b) undoped, (c) Sr-doped, (d) K-doped, (e) Rb-doped, (f) Cs-doped, and (g) Ba vacancy cases and the path projection curves in the $xy$, $yz$, and $xz$ planes.}
    \end{figure*}

We have constructed a spatial coordinate system and obtained the spatial coordinates of H with the FIG. ~\ref{FigS9}. By selecting the spatial coordinates of protons at the initial, final, and intermediate transition states, we have obtained the migration pathway of protons from O$_i$ to O$_f$, as shown in FIG.~\ref{Fig7}. FIG.~\ref{Fig7} illustrates the path curves plotted based on spatial coordinates during the proton migration process, revealing the changes in the relative length of the migration paths. It is evident from FIG.~\ref{Fig7}a that the proton migration path is shortest in undoped BaHfO$_3$, while it is longest in BaHfO$_3$ with V$_Ba$. This observation is further highlighted in the enlarged views in FIG.~\ref{Fig7}(b) and FIG.~\ref{Fig7}(g). In FIG.~\ref{Fig7}(b), the proton migration path trend in Sr-doped BaHfO$_3$ mirrors that of undoped BaHfO$_3$ in the yz plane, with oscillations in the x direction, likely attributed to the slight variance in the radii of Sr$^{2+}$ and Ba$^{2+}$. However, the amplitude of this oscillation is minimal (less than 0.01 coordinate units), having negligible impact on the overall proton migration path. Comparing the enlarged views of the paths of K$^{+}$, Rb$^{+}$, and Cs$^{+}$ doped BaHfO$_3$ in FIG.~\ref{Fig7}(d)-\ref{Fig7}(f), it is evident that the migration trends in the yz-plane for all three dopants closely resemble that of the undoped material. The amplitudes of the oscillations in the x-direction are approximately 0.14, 0.15, and 0.18 coordinate units, respectively. Consequently, the proton migration path in Cs$^{1+}$ doped BaHfO$_3$ emerges as the longest among the three.

This phenomenon is supported in FIG.~\ref{FigS6}, where dopant atoms at the A site near the proton instigate varying degrees of charge interaction, prompting the proton's initial position to migrate toward the doped A site (inclusive of V$_{Ba}$ situations). This interaction results in the bending of the proton's migration path, consequently elongating its trajectory, aligning well with the outcomes observed in FIG.~\ref{Fig7}.

Moreover, given that the yz-plane is either aligned with or closely aligned to the O$_i$-Hf-O$_f$ plane within the crystal, we can approximate the projection of the proton migration pathway on the yz-plane as the trajectory of the proton with the coordinated oxygen. Additionally, the projections on the xy and xz planes can be interpreted as the proton's displacement during the rotation process. These findings strongly support the conclusions illustrated in FIG.~\ref{Fig6}, indicating a high level of agreement regarding the changes in the proton migration trajectory.

\section{Conclusion}

In this study, we have deeply analyzed the effect of A-site defect induction on the proton conduction properties of BaHfO$_3$ through first-principles calculations, aiming to provide theoretical support for the development of electrolyte materials for SOFCs. Our results reveal that BaHfO$_3$ shows great potential as a proton conduction electrolyte with a very low proton migration energy barrier, which suggests that BaHfO$_3$ can realize efficient proton conduction. Through A-site defect induction, especially low-valence ion doping and the introduction of Ba vacancies, we were able to effectively reduce the \( E_{\text{vac}} \) and thus increase the proton concentration. This finding is crucial for improving the proton conductivity of electrolyte materials.

This study sheds light on the proton migration mechanism in BaHfO$_3$, showing that it mainly occurs through the Grotthuss mechanism rather than the vehicle mechanism. This finding enhances our understanding of proton transport behavior in the material and lays a theoretical foundation for designing more efficient proton conductors. By studying the changes in lattice parameters during proton migration, we have determined that doping or controlling vacancies does not change the mode of H$^{+}$ migration but does impact the migration pathway and barrier. This insight is essential for understanding how doping affects the material's structure and performance. Moreover, our research reveals that doping with low-valence ions significantly increases the proton concentration in BaHfO$_3$ without significantly raising the proton migration barrier. This approach proves to be an effective way to enhance the proton conduction properties of BaHfO$_3$. These findings provide novel strategies for optimizing proton conduction properties through A-site defect induction and lay the groundwork for developing new, highly efficient SOFC electrolyte materials. Additionally, we find that the transfer process of protons in the crystal consists of two parts: the migration of protons along with coordinated oxygen and the migration of protons themselves. The radius and valence of A-site ions govern these aspects of proton migration. In conclusion, our research not only establishes a solid theoretical foundation for utilizing BaHfO$_3$ as an H-SOFC electrolyte but also offers fresh perspectives for designing and optimizing other proton conductor materials. Through further experimental research and material development, we can anticipate the creation of more efficient and stable SOFC electrolyte materials, driving the progress of clean energy technology.

\noindent
\underbar{\bf Acknowledgements:}
This work was supported by the Double First-Class Construction Fund for Teacher Development Projects No. 0515024GH0201201 and No. 0515024SH0201201, Shaanxi Provincial Key Research and Development Plan (Grant No. 2024GX-YBXM-456), National Natural Science Foundation of China (Grant No. 12404463 and No. 12474218), and Beijing Natural Science Foundation (Grant No. 1242022).

\noindent
\underbar{\bf Data availability:}
The data that support the findings of this article are openly available\cite{Feng2025}.


\appendix

\begin{figure}[htbp]
\centering
\includegraphics[scale=0.75]{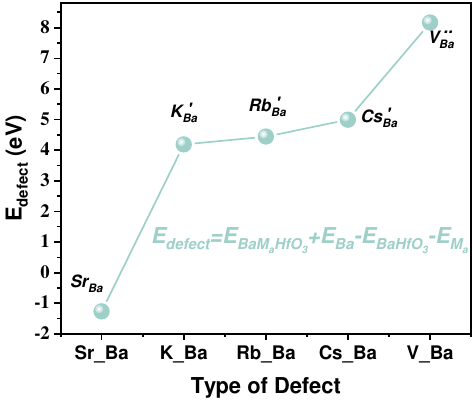}
\caption{\label{FigS1}Formation energy of type of defects.}
\end{figure}

\begin{figure}[htbp]
\centering
\includegraphics[scale=0.65]{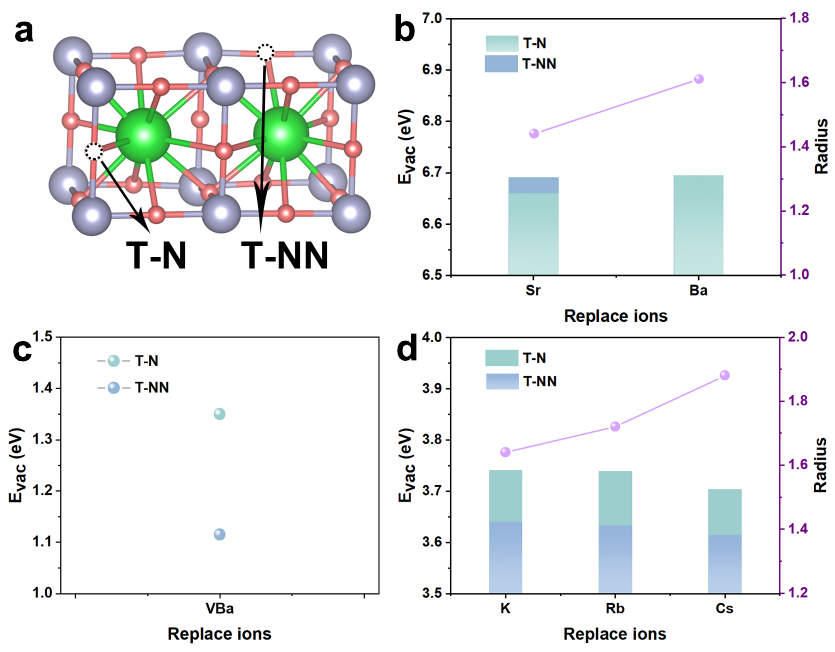}
\caption{\label{FigS2}(a) Schematic of nearest-neighbor versus next-nearest-neighbor oxygen vacancies. (b)-(d) Plots of \( E_{\text{vac}} \) versus Sr$^{2+}$, Ba$^{2+}$, V$_{Ba}$, K$^{+}$, Rb$^{+}$, and Cs$^{+}$ ionic radii for T-N versus T-NN.}
\end{figure}
    
\begin{figure}[htbp]
\centering
\includegraphics[scale=0.86]{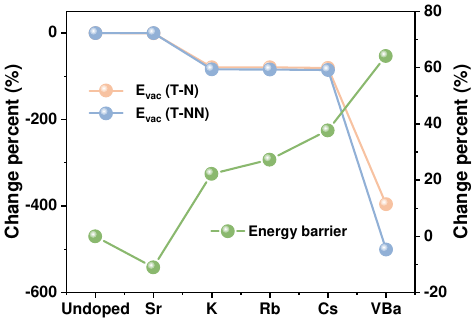}
\caption{\label{FigS3}Magnitude curves of the effect of different A-site modulations on the \( E_{\text{vac}} \) and proton migration energy barrier of BaHfO$_3$.}
\end{figure}

\begin{figure}[htbp]
\centering
\includegraphics[scale=0.45]{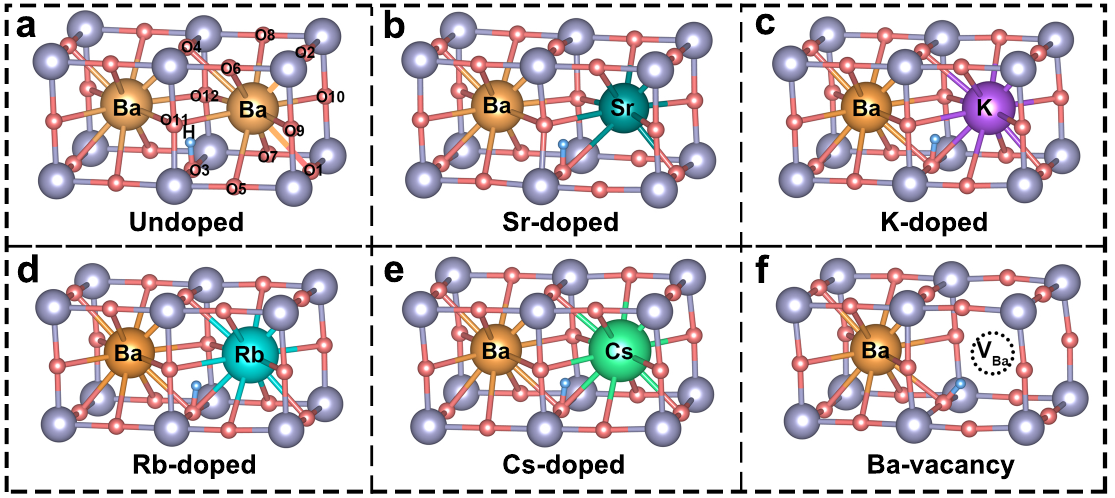}
\caption{\label{FigS4}(a) Undoped, (b) Sr-doped, (c) K-doped, (d) Rb-doped, (c) Cs-doped, and (f) Ba-vacancy lattice distortion diagram.}
\end{figure}
    
\begin{figure}[htbp]
\centering
\includegraphics[scale=0.55]{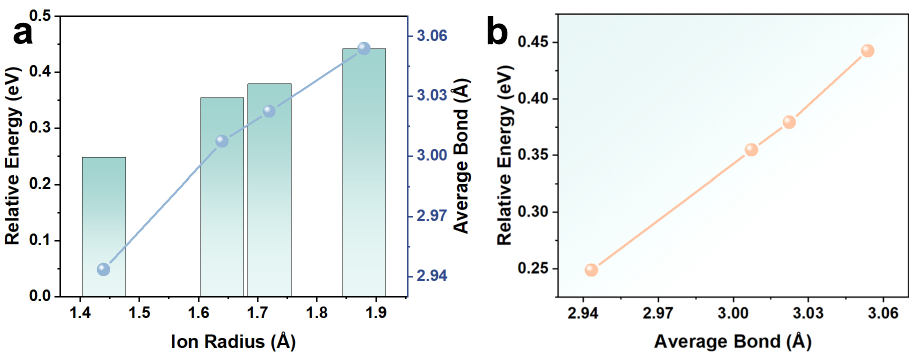}
\caption{\label{FigS5}The relationship between the (a) ionic radius and the migration energy barrier as well as the average bond length of each substituent-coordinate oxygen. (b) Average bond length and migration energy.}
\end{figure}

\begin{figure}[htbp]
\centering
\includegraphics[scale=0.37]{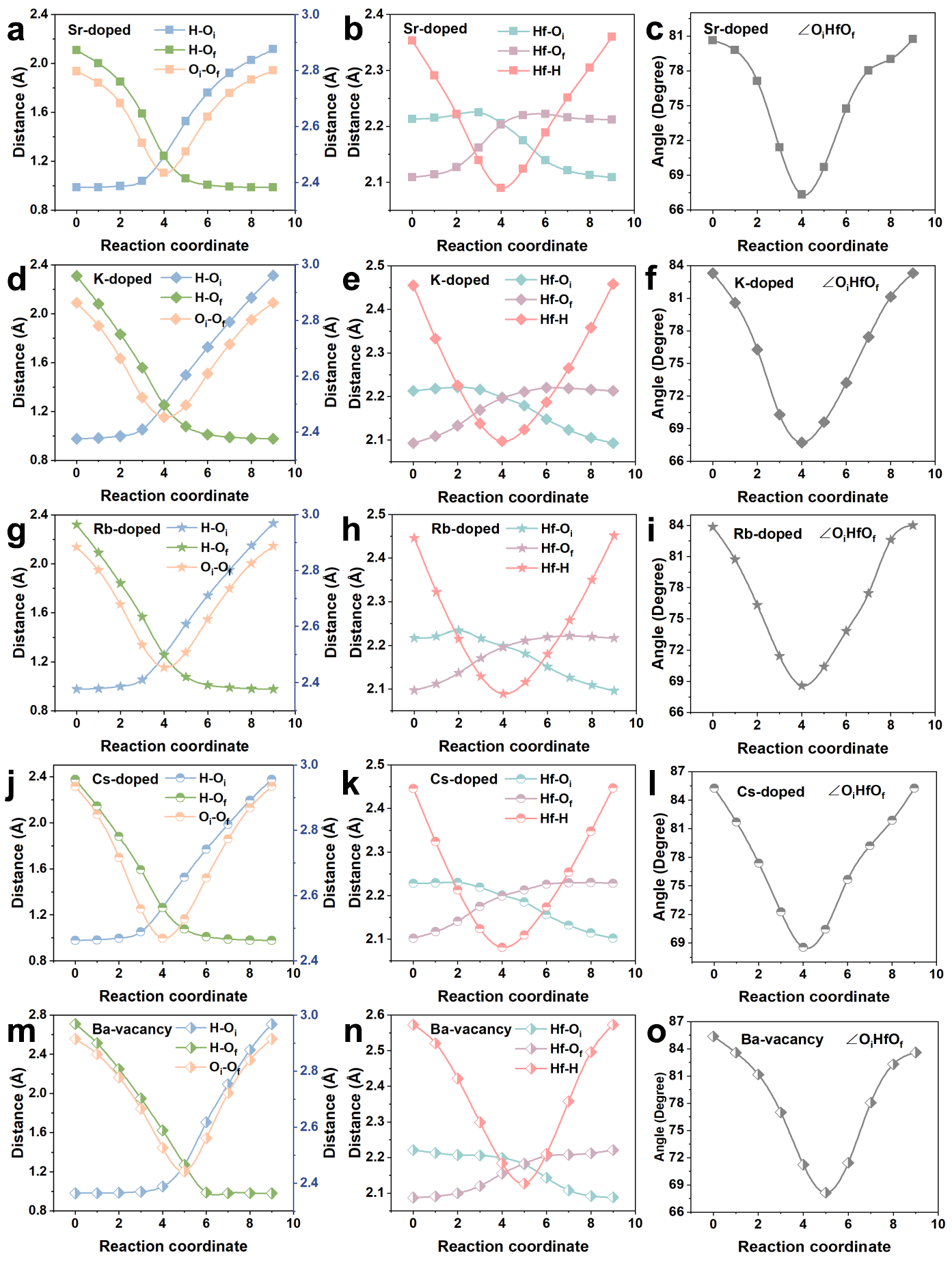}
\caption{\label{FigS6}(a)-(o) Sr, K, Rb, Cs doping and Ba vacancy scenarios during proton migration H-O$_i$, H-O$_f$, O$_i$-O$_f$, Hf-O$_i$, Hf-O$_f$, and Hf-H distances, and angular variation of $\angle$O$_i$HfO$_f$.}
\end{figure}

\begin{figure}[htbp]
\centering
\includegraphics[scale=0.45]{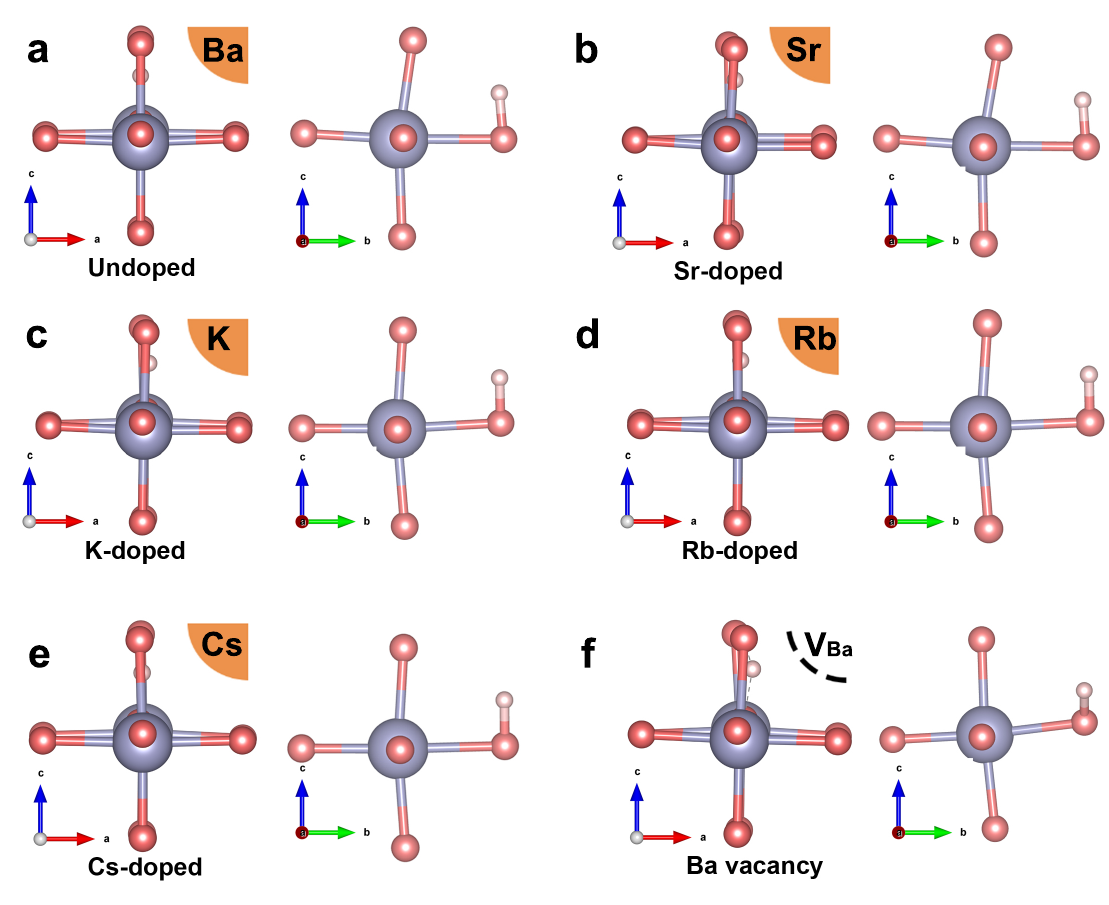}
\caption{\label{FigS7}Changes in HfO$_6$ octahedral and H$^+$ space configurations during proton migration for Sr, K, Rb, Cs doping and Ba vacancy cases.}
\end{figure}

\begin{figure}[htbp]
\centering
\includegraphics[scale=1]{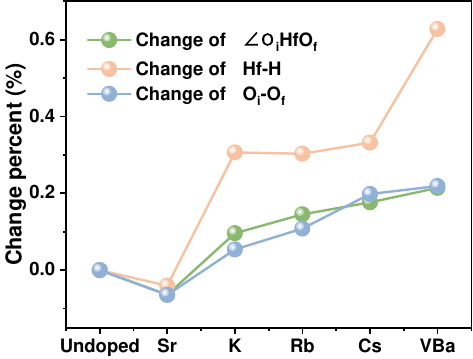}
\caption{\label{FigS8}Degree of change of Hf-H, O$_i$-O$_f$, and $\angle$O$_i$HfO$_f$ compared to undoped BaHfO$_3$.}
\end{figure}

\begin{figure}[htbp]
\centering
\includegraphics[scale=0.8]{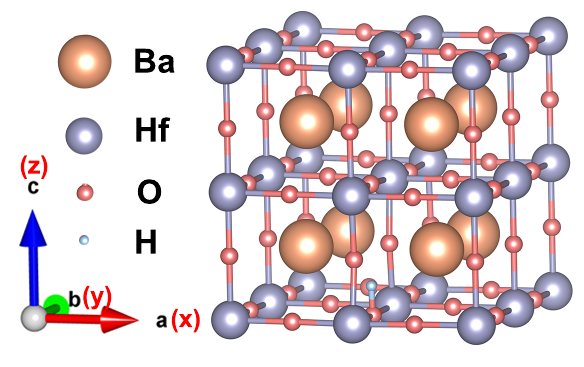}
\caption{\label{FigS9}Axial schematic of a crystalline cell and definition of coordinate axes.}
\end{figure}
    
\section{Appendix}

To thoroughly substantiate the data evidence related to the main text of this paper without compromising the coherence of the narrative, the supporting data have been systematically organized and presented in the form of an appendix.

Figure~\ref{FigS1} shows the formation energy of various defects ($E_\text{defect}$) after the Ba site was replaced. Among them, Sr$_{Ba}$ defects have the lowest formation energy due to the similar valence state and ionic radii of Sr and Ba. Among the three doped monovalent ions, the defect energy, $E_\text{defect}$, gradually increases with the ionic radius. Notably, the Ba vacancy exhibits the highest $E_\text{defect}$ value.

Figure~\ref{FigS2} displays the first-principles calculation outcomes for the formation energies (\( E_{\text{vac}} \)) of oxygen vacancies nearest (T-N) and next nearest (T-NN) to Ba$^{2+}$ in BaHfO$_3$. Schematic diagrams of T-N oxygen vacancies and T-NN oxygen vacancies are depicted in FIG. ~\ref{FigS2}(a). The results indicate that upon doping with Sr$^{2+}$, which is isoelectronic and smaller compared to Ba$^{2+}$, the \( E_{\text{vac}} \) for T-N decreases while that for T-NN remains essentially unchanged (FIG. ~\ref{FigS2}(b)). Upon doping with lower-valence ions, as the radius increases progressively from K$^{+}$ to Cs$^{+}$, 
and E$_{\text{vac}}$ also gradually decreases (FIG. ~\ref{FigS2}(d)). The presence of Ba vacancies (V$_{Ba}^{''}$) significantly reduces E$_{\text{vac}}$ , as shown in FIG. ~\ref{FigS2}(c).
 
Figure~\ref{FigS3} compares the impact of different defect induction processes at the A site on both Evac and the proton migration barrier. It was observed that doping with  Sr$^{2+}$, which is isoelectronic with Ba$^{2+}$, predominantly influences the migration barrier, while doping with lower-valence ions has a more decisive effect on the variation of Evac.

To evaluate the lattice distortion under different A-site defect induction strategies, we conducted an in-depth structural analysis of the material. Figure ~\ref{FigS4} illustrates the changes in the crystal structure, distribution of coordinating oxygen, and the polyhedral environment of the modified site under different A-site defect induction states.

Figure~\ref{FigS5} examines the relationship between ionic radii, bond lengths of modified sites and surrounding coordinating oxygen, and migration energy barriers. It was observed that there is a linear relationship between these pairs, suggesting that changes in lattice parameters due to ionic radii are a direct cause affecting the migration energy barriers.

Figure~\ref{FigS6} illustrates the changes in bond lengths and bond angles during the proton migration from the initial oxygen site O$_i$ to the final oxygen site O$_f$ under doping and Ba vacancy conditions. It is evident that the trends of bond length and bond angle changes are highly consistent across all conditions, indicating that the mode of proton migration remains unchanged regardless of whether the defect induction is through doping or Ba vacancy.

Figure~\ref{FigS7} illustrates the spatial variations of the [HfO$_{6}$] octahedron and proton under different A-site defect induction conditions. It is evident that the ionic state at the modified site, which differs in valence, radius, or vacancy, leads to distinct distortions and rotation states of the surrounding [HfO$_{6}$] octahedra. Consequently, the spatial position of the proton also changes, thereby affecting the migration energy barrier.

Figure~\ref{FigS8} demonstrates the variations in Hf-H, Oi-Of bond lengths, and the $\angle\text{O$_i$HfO$_f$}$ bond angle during the proton migration process under doping and Ba vacancy conditions. It is evident that the smaller radius Sr$^{2+}$, which is isoelectronic with Ba$^{2+}$, has a more pronounced effect on O$_i$-O$_f$ and $\angle\text{O$_i$HfO$_f$}$, while the lower-valence K$^{+}$ with a similar radius predominantly determines the changes in Hf-H. This indicates that ionic radius and valence state each influence different proton migration processes.

We define the x, y and z directions using the a, b and c axes of the $2 \times 2 \times 2$ cell(FIG.~\ref{FigS9}). The xy, xz, and yz planes are created where these directions intersect. Using the $(0,0,0)$ point in the cell as the origin, we have determined the migration pathway of protons from O$_i$ to the O$_f$ by recording the spatial coordinates of protons at the initial, final, and intermediate transition states.

\begin{table}[htbp]
\centering
\caption{Cell parameters of pure BaHfO$_3$.}
\label{Table S1}
\renewcommand{\arraystretch}{1.5}
\begin{tabular}{|c|c|}
\hline
\textbf{Bond} & \textbf{BaHfO$_3$(\AA)}  \\ \hline
Ba-O1 & 2.95589   \\ 
Ba-O2 & 2.94135   \\ 
Ba-O3 & 2.94834   \\ 
\textbf{Average} & 2.94853   \\ \hline
\end{tabular} 
\end{table}

\begin{table*}[htbp]
\centering
\caption{Cell parameters of the polyhedron at the substitutional site in the undoped, doped, and vacancy cases under proton conduction.}
\label{Table S2}
\renewcommand{\arraystretch}{1.5}
\begin{tabular}{|c|c|c|c|c|c|}
\hline
\textbf{Bond} & \textbf{BaHfO$_3$-H (\AA)} & \textbf{BaHfO$_3$-H.K (\AA)} & \textbf{BaHfO$_3$-H.Sr (\AA)} & \textbf{BaHfO$_3$-H.Rb (\AA)} & \textbf{BaHfO$_3$-H.Cs (\AA)} \\ \hline
Ba-O1 & 2.92364 & 2.85487 & 2.93554 & 2.86399 & 2.930192 \\ 
Ba-O2 & 2.82161 & 2.66213 & 2.81053 & 2.88396 & 2.94487 \\ 
Ba-O3 & 3.12231 & 3.24251 & 3.00774 & 3.37005 & 3.30745 \\ 
Ba-O4 & 3.0692  & 3.09153 & 3.07154 & 3.36078 & 3.31112 \\ 
Ba-O5 & 3.2070  & 3.30757 & 3.15257 & 3.12237 & 3.11660 \\ 
Ba-O6 & 3.2070  & 3.29373 & 3.31145 & 3.11385 & 3.11659 \\ 
Ba-O7 & 3.09292 & 2.98979 & 3.11412 & 3.10583 & 3.11566 \\ 
Ba-O8 & 2.99270 & 2.98301 & 3.31271 & 3.07305 & 3.12133 \\ 
Ba-O9 & 3.1070  & 3.11772 & 3.30903 & 3.07330 & 3.12312 \\ 
Ba-O10 & 2.9222 & 3.10593 & 3.12117 & 3.10594 & 3.10661 \\ 
\textbf{Average} & 2.97166 & 2.94336 & 3.00733 & 3.02256 & 3.05378 \\ \hline
\multicolumn{6}{|c|}{} \\ \hline

\multicolumn{2}{|c|}{\textbf{Bond}} & \textbf{BaHfO$_3$-H (\AA)} & \multicolumn{3}{c|}{\textbf{BaHfO$_3$-H.V$_{\text{Ba}}$ (\AA)}} \\ \hline
\multicolumn{2}{|c|}{O1-O4} & 5.98535 & \multicolumn{3}{c|}{6.01584} \\ 
\multicolumn{2}{|c|}{O2-O3} & 5.97715 & \multicolumn{3}{c|}{5.65584} \\ 
\multicolumn{2}{|c|}{O5-O8} & 5.98569 & \multicolumn{3}{c|}{6.16612} \\
\multicolumn{2}{|c|}{O6-O7} & 5.94812 & \multicolumn{3}{c|}{6.16618} \\ 
\multicolumn{2}{|c|}{O9-O12} & 5.87037 & \multicolumn{3}{c|}{6.16669} \\ 
\multicolumn{2}{|c|}{O10-O11} & 5.84731 & \multicolumn{3}{c|}{6.16661} \\ 
\multicolumn{2}{|c|}{\textbf{Average}} & 5.93617 & \multicolumn{3}{c|}{6.05626} \\ \hline
\end{tabular} 
\end{table*}

\begin{table*}[htbp]
\centering
\caption{Cell parameters of the polyhedron at the adjacent site in the undoped, doped and vacancy cases under proton conduction.}
\label{Table S3}
\renewcommand{\arraystretch}{1.5}
\scriptsize
\begin{tabular}{|c|c|c|c|c|c|c|}
\hline
\textbf{Bond} & \textbf{BaHfO$_3$-H (\AA)} & \textbf{BaHfO$_3$-H-Sr (\AA)} & \textbf{BaHfO$_3$-H-K (\AA)} & \textbf{BaHfO$_3$-H-Rb (\AA)} & \textbf{BaHfO$_3$-H-Cs (\AA)} & \textbf{BaHfO$_3$-H-V$_{\text{Ba}}$ (\AA)} \\
\hline
Ba-O1  & 2.92364 & 2.95315 & 2.94622 & 2.9379  & 2.94131 & 3.0311  \\
Ba-O2  & 2.82161 & 2.95315 & 2.85517 & 2.85023 & 2.85096 & 2.93303 \\
Ba-O3  & 3.15574 & 3.15105 & 3.06557 & 3.05927 & 3.05676 & 3.01119 \\
Ba-O4  & 3.0692  & 3.06999 & 2.99873 & 2.99311 & 3.00347 & 2.88932 \\
Ba-O5  & 2.98213 & 2.95995 & 2.95191 & 2.95441 & 2.95721 & 3.0028  \\
Ba-O6  & 2.91882 & 2.9108  & 2.95918 & 2.96034 & 2.96784 & 2.99989 \\
Ba-O7  & 3.08303 & 2.96167 & 2.973  & 3.01781 & 3.00757 & 2.88991 \\
Ba-O8  & 2.90127 & 2.94718 & 2.94532 & 2.93411 & 2.94727 & 2.90207 \\
Ba-O9  & 2.89721 & 2.9291  & 2.93055 & 2.93221 & 2.9298  & 2.91112 \\
Ba-O10 & 2.89672 & 2.927   & 2.93055 & 2.93218 & 2.93428 & 2.92901 \\
Ba-O11 & 2.93072 & 2.94956 & 2.9447  & 2.95184 & 2.95183 & 2.94721 \\
Ba-O12 & 2.97456 & 2.94629 & 3.00352 & 3.02813 & 3.03099 & 2.98871 \\
\hline
\textbf{Average} & \textbf{2.97166} & \textbf{2.97216} & \textbf{2.96791} & \textbf{2.97275} & \textbf{2.98196} & \textbf{2.98086} \\
\hline
\end{tabular} 
\end{table*}

\bibliography{ref.bib}

\begin{thebibliography}{36}%
\makeatletter
\providecommand \@ifxundefined [1]{%
 \@ifx{#1\undefined}
}%
\providecommand \@ifnum [1]{%
 \ifnum #1\expandafter \@firstoftwo
 \else \expandafter \@secondoftwo
 \fi
}%
\providecommand \@ifx [1]{%
 \ifx #1\expandafter \@firstoftwo
 \else \expandafter \@secondoftwo
 \fi
}%
\providecommand \natexlab [1]{#1}%
\providecommand \enquote  [1]{``#1''}%
\providecommand \bibnamefont  [1]{#1}%
\providecommand \bibfnamefont [1]{#1}%
\providecommand \citenamefont [1]{#1}%
\providecommand \href@noop [0]{\@secondoftwo}%
\providecommand \href [0]{\begingroup \@sanitize@url \@href}%
\providecommand \@href[1]{\@@startlink{#1}\@@href}%
\providecommand \@@href[1]{\endgroup#1\@@endlink}%
\providecommand \@sanitize@url [0]{\catcode `\\12\catcode `\$12\catcode
  `\&12\catcode `\#12\catcode `\^12\catcode `\_12\catcode `\%12\relax}%
\providecommand \@@startlink[1]{}%
\providecommand \@@endlink[0]{}%
\providecommand \url  [0]{\begingroup\@sanitize@url \@url }%
\providecommand \@url [1]{\endgroup\@href {#1}{\urlprefix }}%
\providecommand \urlprefix  [0]{URL }%
\providecommand \Eprint [0]{\href }%
\providecommand \doibase [0]{https://doi.org/}%
\providecommand \selectlanguage [0]{\@gobble}%
\providecommand \bibinfo  [0]{\@secondoftwo}%
\providecommand \bibfield  [0]{\@secondoftwo}%
\providecommand \translation [1]{[#1]}%
\providecommand \BibitemOpen [0]{}%
\providecommand \bibitemStop [0]{}%
\providecommand \bibitemNoStop [0]{.\EOS\space}%
\providecommand \EOS [0]{\spacefactor3000\relax}%
\providecommand \BibitemShut  [1]{\csname bibitem#1\endcsname}%
\let\auto@bib@innerbib\@empty
\bibitem [{\citenamefont {Xu}\ \emph {et~al.}(2022)\citenamefont {Xu},
  \citenamefont {Guo}, \citenamefont {Xia}, \citenamefont {He}, \citenamefont
  {Li}, \citenamefont {Bello}, \citenamefont {Zheng},\ and\ \citenamefont
  {Ni}}]{xu2022comprehensive}%
  \BibitemOpen
  \bibfield  {author} {\bibinfo {author} {\bibfnamefont {Q.}~\bibnamefont
  {Xu}}, \bibinfo {author} {\bibfnamefont {Z.}~\bibnamefont {Guo}}, \bibinfo
  {author} {\bibfnamefont {L.}~\bibnamefont {Xia}}, \bibinfo {author}
  {\bibfnamefont {Q.}~\bibnamefont {He}}, \bibinfo {author} {\bibfnamefont
  {Z.}~\bibnamefont {Li}}, \bibinfo {author} {\bibfnamefont {I.~T.}\
  \bibnamefont {Bello}}, \bibinfo {author} {\bibfnamefont {K.}~\bibnamefont
  {Zheng}},\ and\ \bibinfo {author} {\bibfnamefont {M.}~\bibnamefont {Ni}},\
  }\bibfield  {title} {\bibinfo {title} {A comprehensive review of solid oxide
  fuel cells operating on various promising alternative fuels},\ }\href
  {https://doi.org/10.1016/j.enconman.2021.115175} {\bibfield  {journal}
  {\bibinfo  {journal} {Energy Conversion and Management}\ }\textbf {\bibinfo
  {volume} {253}},\ \bibinfo {pages} {115175} (\bibinfo {year}
  {2022})}\BibitemShut {NoStop}%
\bibitem [{\citenamefont {Bicer}\ and\ \citenamefont
  {Khalid}(2020)}]{bicer2020life}%
  \BibitemOpen
  \bibfield  {author} {\bibinfo {author} {\bibfnamefont {Y.}~\bibnamefont
  {Bicer}}\ and\ \bibinfo {author} {\bibfnamefont {F.}~\bibnamefont {Khalid}},\
  }\bibfield  {title} {\bibinfo {title} {Life cycle environmental impact
  comparison of solid oxide fuel cells fueled by natural gas, hydrogen, ammonia
  and methanol for combined heat and power generation},\ }\href
  {https://doi.org/10.1016/j.ijhydene.2018.11.122} {\bibfield  {journal}
  {\bibinfo  {journal} {International Journal of Hydrogen Energy}\ }\textbf
  {\bibinfo {volume} {45}},\ \bibinfo {pages} {3670} (\bibinfo {year}
  {2020})}\BibitemShut {NoStop}%
\bibitem [{\citenamefont {Timurkutluk}\ \emph {et~al.}(2016)\citenamefont
  {Timurkutluk}, \citenamefont {Timurkutluk}, \citenamefont {Mat},\ and\
  \citenamefont {Kaplan}}]{timurkutluk2016review}%
  \BibitemOpen
  \bibfield  {author} {\bibinfo {author} {\bibfnamefont {B.}~\bibnamefont
  {Timurkutluk}}, \bibinfo {author} {\bibfnamefont {C.}~\bibnamefont
  {Timurkutluk}}, \bibinfo {author} {\bibfnamefont {M.~D.}\ \bibnamefont
  {Mat}},\ and\ \bibinfo {author} {\bibfnamefont {Y.}~\bibnamefont {Kaplan}},\
  }\bibfield  {title} {\bibinfo {title} {A review on cell/stack designs for
  high performance solid oxide fuel cells},\ }\href
  {https://doi.org/10.1016/j.rser.2015.12.034} {\bibfield  {journal} {\bibinfo
  {journal} {Renewable and Sustainable Energy Reviews}\ }\textbf {\bibinfo
  {volume} {56}},\ \bibinfo {pages} {1101} (\bibinfo {year}
  {2016})}\BibitemShut {NoStop}%
\bibitem [{\citenamefont {Ryu}\ \emph {et~al.}(2023)\citenamefont {Ryu},
  \citenamefont {Choi}, \citenamefont {Kim}, \citenamefont {Lee}, \citenamefont
  {Jeong}, \citenamefont {Yu}, \citenamefont {Cho},\ and\ \citenamefont
  {Cha}}]{ryu2023nanocrystal}%
  \BibitemOpen
  \bibfield  {author} {\bibinfo {author} {\bibfnamefont {S.}~\bibnamefont
  {Ryu}}, \bibinfo {author} {\bibfnamefont {I.~W.}\ \bibnamefont {Choi}},
  \bibinfo {author} {\bibfnamefont {Y.~J.}\ \bibnamefont {Kim}}, \bibinfo
  {author} {\bibfnamefont {S.}~\bibnamefont {Lee}}, \bibinfo {author}
  {\bibfnamefont {W.}~\bibnamefont {Jeong}}, \bibinfo {author} {\bibfnamefont
  {W.}~\bibnamefont {Yu}}, \bibinfo {author} {\bibfnamefont {G.~Y.}\
  \bibnamefont {Cho}},\ and\ \bibinfo {author} {\bibfnamefont {S.~W.}\
  \bibnamefont {Cha}},\ }\bibfield  {title} {\bibinfo {title} {Nanocrystal
  engineering of thin-film yttria-stabilized zirconia electrolytes for
  low-temperature solid-oxide fuel cells},\ }\href
  {https://doi.org/10.1021/acsami.3c09025} {\bibfield  {journal} {\bibinfo
  {journal} {ACS applied materials \& interfaces}\ }\textbf {\bibinfo {volume}
  {15}},\ \bibinfo {pages} {42659} (\bibinfo {year} {2023})}\BibitemShut
  {NoStop}%
\bibitem [{\citenamefont {Ding}\ \emph {et~al.}(2020)\citenamefont {Ding},
  \citenamefont {Wu}, \citenamefont {Jiang}, \citenamefont {Ding},
  \citenamefont {Bian}, \citenamefont {Hu}, \citenamefont {Singh},
  \citenamefont {Orme}, \citenamefont {Wang}, \citenamefont {Zhang} \emph
  {et~al.}}]{ding2020self}%
  \BibitemOpen
  \bibfield  {author} {\bibinfo {author} {\bibfnamefont {H.}~\bibnamefont
  {Ding}}, \bibinfo {author} {\bibfnamefont {W.}~\bibnamefont {Wu}}, \bibinfo
  {author} {\bibfnamefont {C.}~\bibnamefont {Jiang}}, \bibinfo {author}
  {\bibfnamefont {Y.}~\bibnamefont {Ding}}, \bibinfo {author} {\bibfnamefont
  {W.}~\bibnamefont {Bian}}, \bibinfo {author} {\bibfnamefont {B.}~\bibnamefont
  {Hu}}, \bibinfo {author} {\bibfnamefont {P.}~\bibnamefont {Singh}}, \bibinfo
  {author} {\bibfnamefont {C.~J.}\ \bibnamefont {Orme}}, \bibinfo {author}
  {\bibfnamefont {L.}~\bibnamefont {Wang}}, \bibinfo {author} {\bibfnamefont
  {Y.}~\bibnamefont {Zhang}}, \emph {et~al.},\ }\bibfield  {title} {\bibinfo
  {title} {Self-sustainable protonic ceramic electrochemical cells using a
  triple conducting electrode for hydrogen and power production},\ }\href
  {https://doi.org/10.1038/s41467-020-15677-z} {\bibfield  {journal} {\bibinfo
  {journal} {Nature communications}\ }\textbf {\bibinfo {volume} {11}},\
  \bibinfo {pages} {1907} (\bibinfo {year} {2020})}\BibitemShut {NoStop}%
\bibitem [{\citenamefont {Saito}\ and\ \citenamefont
  {Yashima}(2023)}]{saito2023high}%
  \BibitemOpen
  \bibfield  {author} {\bibinfo {author} {\bibfnamefont {K.}~\bibnamefont
  {Saito}}\ and\ \bibinfo {author} {\bibfnamefont {M.}~\bibnamefont
  {Yashima}},\ }\bibfield  {title} {\bibinfo {title} {High proton conductivity
  within the ‘norby gap’by stabilizing a perovskite with disordered
  intrinsic oxygen vacancies},\ }\href
  {https://doi.org/10.1038/s41467-023-43122-4} {\bibfield  {journal} {\bibinfo
  {journal} {Nature Communications}\ }\textbf {\bibinfo {volume} {14}},\
  \bibinfo {pages} {7466} (\bibinfo {year} {2023})}\BibitemShut {NoStop}%
\bibitem [{\citenamefont {Lu}\ \emph {et~al.}(2022)\citenamefont {Lu},
  \citenamefont {Zhang}, \citenamefont {Wang}, \citenamefont {Li},
  \citenamefont {Qiao}, \citenamefont {Zhao}, \citenamefont {He}, \citenamefont
  {Lu}, \citenamefont {Li}, \citenamefont {Wu} \emph
  {et~al.}}]{lu2022enhanced}%
  \BibitemOpen
  \bibfield  {author} {\bibinfo {author} {\bibfnamefont {N.}~\bibnamefont
  {Lu}}, \bibinfo {author} {\bibfnamefont {Z.}~\bibnamefont {Zhang}}, \bibinfo
  {author} {\bibfnamefont {Y.}~\bibnamefont {Wang}}, \bibinfo {author}
  {\bibfnamefont {H.-B.}\ \bibnamefont {Li}}, \bibinfo {author} {\bibfnamefont
  {S.}~\bibnamefont {Qiao}}, \bibinfo {author} {\bibfnamefont {B.}~\bibnamefont
  {Zhao}}, \bibinfo {author} {\bibfnamefont {Q.}~\bibnamefont {He}}, \bibinfo
  {author} {\bibfnamefont {S.}~\bibnamefont {Lu}}, \bibinfo {author}
  {\bibfnamefont {C.}~\bibnamefont {Li}}, \bibinfo {author} {\bibfnamefont
  {Y.}~\bibnamefont {Wu}}, \emph {et~al.},\ }\bibfield  {title} {\bibinfo
  {title} {Enhanced low-temperature proton conductivity in
  hydrogen-intercalated brownmillerite oxide},\ }\href
  {https://doi.org/10.1038/s41560-022-01166-8} {\bibfield  {journal} {\bibinfo
  {journal} {Nature Energy}\ }\textbf {\bibinfo {volume} {7}},\ \bibinfo
  {pages} {1208} (\bibinfo {year} {2022})}\BibitemShut {NoStop}%
\bibitem [{\citenamefont {Islam}\ \emph {et~al.}(2020)\citenamefont {Islam},
  \citenamefont {Nolan}, \citenamefont {Wang}, \citenamefont {Bai},\ and\
  \citenamefont {Mo}}]{islam2020computational}%
  \BibitemOpen
  \bibfield  {author} {\bibinfo {author} {\bibfnamefont {M.~S.}\ \bibnamefont
  {Islam}}, \bibinfo {author} {\bibfnamefont {A.~M.}\ \bibnamefont {Nolan}},
  \bibinfo {author} {\bibfnamefont {S.}~\bibnamefont {Wang}}, \bibinfo {author}
  {\bibfnamefont {Q.}~\bibnamefont {Bai}},\ and\ \bibinfo {author}
  {\bibfnamefont {Y.}~\bibnamefont {Mo}},\ }\bibfield  {title} {\bibinfo
  {title} {A computational study of fast proton diffusion in brownmillerite
  sr2co2o5},\ }\href {https://doi.org/10.1021/acs.chemmater.0c00544} {\bibfield
   {journal} {\bibinfo  {journal} {Chemistry of Materials}\ }\textbf {\bibinfo
  {volume} {32}},\ \bibinfo {pages} {5028} (\bibinfo {year}
  {2020})}\BibitemShut {NoStop}%
\bibitem [{\citenamefont {Ozawa}\ \emph {et~al.}(2018)\citenamefont {Ozawa},
  \citenamefont {Ogihara}, \citenamefont {Hasegawa}, \citenamefont {Hiruta},
  \citenamefont {Ohba},\ and\ \citenamefont {Kishida}}]{ozawa2018intercalated}%
  \BibitemOpen
  \bibfield  {author} {\bibinfo {author} {\bibfnamefont {Y.}~\bibnamefont
  {Ozawa}}, \bibinfo {author} {\bibfnamefont {N.}~\bibnamefont {Ogihara}},
  \bibinfo {author} {\bibfnamefont {M.}~\bibnamefont {Hasegawa}}, \bibinfo
  {author} {\bibfnamefont {O.}~\bibnamefont {Hiruta}}, \bibinfo {author}
  {\bibfnamefont {N.}~\bibnamefont {Ohba}},\ and\ \bibinfo {author}
  {\bibfnamefont {Y.}~\bibnamefont {Kishida}},\ }\bibfield  {title} {\bibinfo
  {title} {Intercalated metal--organic frameworks with high electronic
  conductivity as negative electrode materials for hybrid capacitors},\ }\href
  {https://doi.org/10.1038/s42004-018-0064-5} {\bibfield  {journal} {\bibinfo
  {journal} {Communications Chemistry}\ }\textbf {\bibinfo {volume} {1}},\
  \bibinfo {pages} {65} (\bibinfo {year} {2018})}\BibitemShut {NoStop}%
\bibitem [{\citenamefont {Sun}\ \emph {et~al.}(2014)\citenamefont {Sun},
  \citenamefont {Shi}, \citenamefont {Liu}, \citenamefont {Bi},\ and\
  \citenamefont {Liu}}]{sun2014easily}%
  \BibitemOpen
  \bibfield  {author} {\bibinfo {author} {\bibfnamefont {W.}~\bibnamefont
  {Sun}}, \bibinfo {author} {\bibfnamefont {Z.}~\bibnamefont {Shi}}, \bibinfo
  {author} {\bibfnamefont {M.}~\bibnamefont {Liu}}, \bibinfo {author}
  {\bibfnamefont {L.}~\bibnamefont {Bi}},\ and\ \bibinfo {author}
  {\bibfnamefont {W.}~\bibnamefont {Liu}},\ }\bibfield  {title} {\bibinfo
  {title} {An easily sintered, chemically stable, barium zirconate-based proton
  conductor for high-performance proton-conducting solid oxide fuel cells},\
  }\href {https://doi.org/10.1002/adfm.201401478} {\bibfield  {journal}
  {\bibinfo  {journal} {Advanced Functional Materials}\ }\textbf {\bibinfo
  {volume} {24}},\ \bibinfo {pages} {5695} (\bibinfo {year}
  {2014})}\BibitemShut {NoStop}%
\bibitem [{\citenamefont {Gong}\ \emph {et~al.}(2018)\citenamefont {Gong},
  \citenamefont {Sun}, \citenamefont {Jin}, \citenamefont {Miao},\ and\
  \citenamefont {Liu}}]{gong2018barium}%
  \BibitemOpen
  \bibfield  {author} {\bibinfo {author} {\bibfnamefont {Z.}~\bibnamefont
  {Gong}}, \bibinfo {author} {\bibfnamefont {W.}~\bibnamefont {Sun}}, \bibinfo
  {author} {\bibfnamefont {Z.}~\bibnamefont {Jin}}, \bibinfo {author}
  {\bibfnamefont {L.}~\bibnamefont {Miao}},\ and\ \bibinfo {author}
  {\bibfnamefont {W.}~\bibnamefont {Liu}},\ }\bibfield  {title} {\bibinfo
  {title} {Barium-and strontium-containing anode materials toward ceria-based
  solid oxide fuel cells with high open circuit voltages},\ }\href
  {https://doi.org/10.1021/acsaem.8b00825} {\bibfield  {journal} {\bibinfo
  {journal} {ACS Applied Energy Materials}\ }\textbf {\bibinfo {volume} {1}},\
  \bibinfo {pages} {3521} (\bibinfo {year} {2018})}\BibitemShut {NoStop}%
\bibitem [{\citenamefont {Lyagaeva}\ \emph {et~al.}(2016)\citenamefont
  {Lyagaeva}, \citenamefont {Danilov}, \citenamefont {Vdovin}, \citenamefont
  {Bu}, \citenamefont {Medvedev}, \citenamefont {Demin},\ and\ \citenamefont
  {Tsiakaras}}]{lyagaeva2016new}%
  \BibitemOpen
  \bibfield  {author} {\bibinfo {author} {\bibfnamefont {J.}~\bibnamefont
  {Lyagaeva}}, \bibinfo {author} {\bibfnamefont {N.}~\bibnamefont {Danilov}},
  \bibinfo {author} {\bibfnamefont {G.}~\bibnamefont {Vdovin}}, \bibinfo
  {author} {\bibfnamefont {J.}~\bibnamefont {Bu}}, \bibinfo {author}
  {\bibfnamefont {D.}~\bibnamefont {Medvedev}}, \bibinfo {author}
  {\bibfnamefont {A.}~\bibnamefont {Demin}},\ and\ \bibinfo {author}
  {\bibfnamefont {P.}~\bibnamefont {Tsiakaras}},\ }\bibfield  {title} {\bibinfo
  {title} {A new dy-doped baceo 3--bazro 3 proton-conducting material as a
  promising electrolyte for reversible solid oxide fuel cells},\ }\href
  {https://doi.org/10.1039/C6TA06414K} {\bibfield  {journal} {\bibinfo
  {journal} {Journal of Materials Chemistry A}\ }\textbf {\bibinfo {volume}
  {4}},\ \bibinfo {pages} {15390} (\bibinfo {year} {2016})}\BibitemShut
  {NoStop}%
\bibitem [{\citenamefont {Kannan}\ \emph {et~al.}(2013)\citenamefont {Kannan},
  \citenamefont {Singh}, \citenamefont {Gill}, \citenamefont
  {F{\"u}rstenhaupt},\ and\ \citenamefont
  {Thangadurai}}]{kannan2013chemically}%
  \BibitemOpen
  \bibfield  {author} {\bibinfo {author} {\bibfnamefont {R.}~\bibnamefont
  {Kannan}}, \bibinfo {author} {\bibfnamefont {K.}~\bibnamefont {Singh}},
  \bibinfo {author} {\bibfnamefont {S.}~\bibnamefont {Gill}}, \bibinfo {author}
  {\bibfnamefont {T.}~\bibnamefont {F{\"u}rstenhaupt}},\ and\ \bibinfo {author}
  {\bibfnamefont {V.}~\bibnamefont {Thangadurai}},\ }\bibfield  {title}
  {\bibinfo {title} {Chemically stable proton conducting doped baceo3-no more
  fear to sofc wastes},\ }\href {https://doi.org/10.1038/srep02138} {\bibfield
  {journal} {\bibinfo  {journal} {Scientific Reports}\ }\textbf {\bibinfo
  {volume} {3}},\ \bibinfo {pages} {2138} (\bibinfo {year} {2013})}\BibitemShut
  {NoStop}%
\bibitem [{\citenamefont {Han}\ and\ \citenamefont {Uda}(2018)}]{han2018best}%
  \BibitemOpen
  \bibfield  {author} {\bibinfo {author} {\bibfnamefont {D.}~\bibnamefont
  {Han}}\ and\ \bibinfo {author} {\bibfnamefont {T.}~\bibnamefont {Uda}},\
  }\bibfield  {title} {\bibinfo {title} {The best composition of an y-doped
  bazro 3 electrolyte: selection criteria from transport properties,
  microstructure, and phase behavior},\ }\href
  {https://doi.org/10.1039/C8TA06280C} {\bibfield  {journal} {\bibinfo
  {journal} {Journal of Materials Chemistry A}\ }\textbf {\bibinfo {volume}
  {6}},\ \bibinfo {pages} {18571} (\bibinfo {year} {2018})}\BibitemShut
  {NoStop}%
\bibitem [{\citenamefont {Hossain}\ \emph {et~al.}(2017)\citenamefont
  {Hossain}, \citenamefont {Abdalla}, \citenamefont {Jamain}, \citenamefont
  {Zaini},\ and\ \citenamefont {Azad}}]{hossain2017review}%
  \BibitemOpen
  \bibfield  {author} {\bibinfo {author} {\bibfnamefont {S.}~\bibnamefont
  {Hossain}}, \bibinfo {author} {\bibfnamefont {A.~M.}\ \bibnamefont
  {Abdalla}}, \bibinfo {author} {\bibfnamefont {S.~N.~B.}\ \bibnamefont
  {Jamain}}, \bibinfo {author} {\bibfnamefont {J.~H.}\ \bibnamefont {Zaini}},\
  and\ \bibinfo {author} {\bibfnamefont {A.~K.}\ \bibnamefont {Azad}},\
  }\bibfield  {title} {\bibinfo {title} {A review on proton conducting
  electrolytes for clean energy and intermediate temperature-solid oxide fuel
  cells},\ }\href {https://doi.org/10.1016/j.rser.2017.05.147} {\bibfield
  {journal} {\bibinfo  {journal} {Renewable and Sustainable Energy Reviews}\
  }\textbf {\bibinfo {volume} {79}},\ \bibinfo {pages} {750} (\bibinfo {year}
  {2017})}\BibitemShut {NoStop}%
\bibitem [{\citenamefont {Bandura}\ \emph {et~al.}(2010)\citenamefont
  {Bandura}, \citenamefont {Evarestov},\ and\ \citenamefont
  {Kuruch}}]{bandura2010hybrid}%
  \BibitemOpen
  \bibfield  {author} {\bibinfo {author} {\bibfnamefont {A.}~\bibnamefont
  {Bandura}}, \bibinfo {author} {\bibfnamefont {R.}~\bibnamefont {Evarestov}},\
  and\ \bibinfo {author} {\bibfnamefont {D.}~\bibnamefont {Kuruch}},\
  }\bibfield  {title} {\bibinfo {title} {Hybrid hf--dft modeling of monolayer
  water adsorption on (001) surface of cubic bahfo3 and bazro3 crystals},\
  }\href {https://doi.org/10.1016/j.susc.2010.05.030} {\bibfield  {journal}
  {\bibinfo  {journal} {Surface science}\ }\textbf {\bibinfo {volume} {604}},\
  \bibinfo {pages} {1591} (\bibinfo {year} {2010})}\BibitemShut {NoStop}%
\bibitem [{\citenamefont {Kang}\ and\ \citenamefont
  {Sholl}(2017)}]{kang2017characterizing}%
  \BibitemOpen
  \bibfield  {author} {\bibinfo {author} {\bibfnamefont {S.~G.}\ \bibnamefont
  {Kang}}\ and\ \bibinfo {author} {\bibfnamefont {D.~S.}\ \bibnamefont
  {Sholl}},\ }\bibfield  {title} {\bibinfo {title} {Characterizing chemical
  stability and proton conductivity of b-site doped barium hafnate (bahfo3) and
  barium stannate (basno3) with first principles modeling},\ }\href
  {https://doi.org/10.1016/j.jallcom.2016.09.221} {\bibfield  {journal}
  {\bibinfo  {journal} {Journal of Alloys and Compounds}\ }\textbf {\bibinfo
  {volume} {693}},\ \bibinfo {pages} {738} (\bibinfo {year}
  {2017})}\BibitemShut {NoStop}%
\bibitem [{\citenamefont {Luo}\ \emph {et~al.}(2024)\citenamefont {Luo},
  \citenamefont {Hu}, \citenamefont {Zhou}, \citenamefont {Ding}, \citenamefont
  {Zhang}, \citenamefont {Li},\ and\ \citenamefont {Liu}}]{luo2024harnessing}%
  \BibitemOpen
  \bibfield  {author} {\bibinfo {author} {\bibfnamefont {Z.}~\bibnamefont
  {Luo}}, \bibinfo {author} {\bibfnamefont {X.}~\bibnamefont {Hu}}, \bibinfo
  {author} {\bibfnamefont {Y.}~\bibnamefont {Zhou}}, \bibinfo {author}
  {\bibfnamefont {Y.}~\bibnamefont {Ding}}, \bibinfo {author} {\bibfnamefont
  {W.}~\bibnamefont {Zhang}}, \bibinfo {author} {\bibfnamefont
  {T.}~\bibnamefont {Li}},\ and\ \bibinfo {author} {\bibfnamefont
  {M.}~\bibnamefont {Liu}},\ }\bibfield  {title} {\bibinfo {title} {Harnessing
  high-throughput computational methods to accelerate the discovery of optimal
  proton conductors for high-performance and durable protonic ceramic
  electrochemical cells},\ }\href {https://doi.org/10.1002/adma.202311159}
  {\bibfield  {journal} {\bibinfo  {journal} {Advanced Materials}\ }\textbf
  {\bibinfo {volume} {36}},\ \bibinfo {pages} {2311159} (\bibinfo {year}
  {2024})}\BibitemShut {NoStop}%
\bibitem [{\citenamefont {Gomez}\ \emph {et~al.}(2012)\citenamefont {Gomez},
  \citenamefont {Shepardson}, \citenamefont {Nguyen},\ and\ \citenamefont
  {Kehinde}}]{gomez2012periodic}%
  \BibitemOpen
  \bibfield  {author} {\bibinfo {author} {\bibfnamefont {M.~A.}\ \bibnamefont
  {Gomez}}, \bibinfo {author} {\bibfnamefont {D.}~\bibnamefont {Shepardson}},
  \bibinfo {author} {\bibfnamefont {L.~T.}\ \bibnamefont {Nguyen}},\ and\
  \bibinfo {author} {\bibfnamefont {T.}~\bibnamefont {Kehinde}},\ }\bibfield
  {title} {\bibinfo {title} {Periodic long range proton conduction pathways in
  pseudo-cubic and orthorhombic perovskites},\ }\href
  {https://doi.org/10.1016/j.ssi.2011.08.001} {\bibfield  {journal} {\bibinfo
  {journal} {Solid State Ionics}\ }\textbf {\bibinfo {volume} {213}},\ \bibinfo
  {pages} {8} (\bibinfo {year} {2012})}\BibitemShut {NoStop}%
\bibitem [{\citenamefont {Bu}\ \emph {et~al.}(2016)\citenamefont {Bu},
  \citenamefont {J{\"o}nsson},\ and\ \citenamefont {Zhao}}]{bu2016effect}%
  \BibitemOpen
  \bibfield  {author} {\bibinfo {author} {\bibfnamefont {J.}~\bibnamefont
  {Bu}}, \bibinfo {author} {\bibfnamefont {P.~G.}\ \bibnamefont
  {J{\"o}nsson}},\ and\ \bibinfo {author} {\bibfnamefont {Z.}~\bibnamefont
  {Zhao}},\ }\bibfield  {title} {\bibinfo {title} {The effect of nio on the
  conductivity of bazr 0.5 ce 0.3 y 0.2 o 3- $\delta$ based electrolytes},\
  }\href {https://doi.org/10.1039/C6RA09936J} {\bibfield  {journal} {\bibinfo
  {journal} {RSC Advances}\ }\textbf {\bibinfo {volume} {6}},\ \bibinfo {pages}
  {62368} (\bibinfo {year} {2016})}\BibitemShut {NoStop}%
\bibitem [{\citenamefont {Guo}\ \emph {et~al.}(2022)\citenamefont {Guo},
  \citenamefont {Li}, \citenamefont {Guan}, \citenamefont {Kong}, \citenamefont
  {Cui}, \citenamefont {Zhou},\ and\ \citenamefont {He}}]{guo2022sn}%
  \BibitemOpen
  \bibfield  {author} {\bibinfo {author} {\bibfnamefont {R.}~\bibnamefont
  {Guo}}, \bibinfo {author} {\bibfnamefont {D.}~\bibnamefont {Li}}, \bibinfo
  {author} {\bibfnamefont {R.}~\bibnamefont {Guan}}, \bibinfo {author}
  {\bibfnamefont {D.}~\bibnamefont {Kong}}, \bibinfo {author} {\bibfnamefont
  {Z.}~\bibnamefont {Cui}}, \bibinfo {author} {\bibfnamefont {Z.}~\bibnamefont
  {Zhou}},\ and\ \bibinfo {author} {\bibfnamefont {T.}~\bibnamefont {He}},\
  }\bibfield  {title} {\bibinfo {title} {Sn--dy--cu triply doped bazr0. 1ce0.
  7y0. 2o3- $\delta$: a chemically stable and highly proton-conductive
  electrolyte for low-temperature solid oxide fuel cells},\ }\href@noop {}
  {\bibfield  {journal} {\bibinfo  {journal} {ACS Sustainable Chemistry \&
  Engineering}\ }\textbf {\bibinfo {volume} {10}},\ \bibinfo {pages} {5352}
  (\bibinfo {year} {2022})}\BibitemShut {NoStop}%
\bibitem [{\citenamefont {Oh}\ \emph {et~al.}(2024)\citenamefont {Oh},
  \citenamefont {Kim}, \citenamefont {Ryu},\ and\ \citenamefont
  {Lee}}]{oh2024novel}%
  \BibitemOpen
  \bibfield  {author} {\bibinfo {author} {\bibfnamefont {S.}~\bibnamefont
  {Oh}}, \bibinfo {author} {\bibfnamefont {D.}~\bibnamefont {Kim}}, \bibinfo
  {author} {\bibfnamefont {H.~J.}\ \bibnamefont {Ryu}},\ and\ \bibinfo {author}
  {\bibfnamefont {K.~T.}\ \bibnamefont {Lee}},\ }\bibfield  {title} {\bibinfo
  {title} {A novel high-entropy perovskite electrolyte with improved proton
  conductivity and stability for reversible protonic ceramic electrochemical
  cells},\ }\href {https://doi.org/10.1002/adfm.202311426} {\bibfield
  {journal} {\bibinfo  {journal} {Advanced Functional Materials}\ }\textbf
  {\bibinfo {volume} {34}},\ \bibinfo {pages} {2311426} (\bibinfo {year}
  {2024})}\BibitemShut {NoStop}%
\bibitem [{\citenamefont {Wang}\ \emph {et~al.}(2024)\citenamefont {Wang},
  \citenamefont {Zheng}, \citenamefont {Sun}, \citenamefont {Zhang},
  \citenamefont {Guo}, \citenamefont {Hu},\ and\ \citenamefont
  {Feng}}]{wang2024large}%
  \BibitemOpen
  \bibfield  {author} {\bibinfo {author} {\bibfnamefont {D.}~\bibnamefont
  {Wang}}, \bibinfo {author} {\bibfnamefont {T.}~\bibnamefont {Zheng}},
  \bibinfo {author} {\bibfnamefont {H.}~\bibnamefont {Sun}}, \bibinfo {author}
  {\bibfnamefont {X.}~\bibnamefont {Zhang}}, \bibinfo {author} {\bibfnamefont
  {X.}~\bibnamefont {Guo}}, \bibinfo {author} {\bibfnamefont {Q.}~\bibnamefont
  {Hu}},\ and\ \bibinfo {author} {\bibfnamefont {Y.}~\bibnamefont {Feng}},\
  }\bibfield  {title} {\bibinfo {title} {Large grain sized and high grain
  boundary conductive bazr0. 1ce0. 7y0. 2o3-$\delta$ (bzcy) proton-conducting
  electrolytes for solid oxide fuel cells by cu doping},\ }\href
  {https://doi.org/10.1016/j.ijhydene.2024.05.270} {\bibfield  {journal}
  {\bibinfo  {journal} {International Journal of Hydrogen Energy}\ }\textbf
  {\bibinfo {volume} {71}},\ \bibinfo {pages} {357} (\bibinfo {year}
  {2024})}\BibitemShut {NoStop}%
\bibitem [{\citenamefont {Ivanova}\ \emph {et~al.}(2012)\citenamefont
  {Ivanova}, \citenamefont {Ricote}, \citenamefont {Meulenberg}, \citenamefont
  {Haugsrud},\ and\ \citenamefont {Ziegner}}]{ivanova2012effects}%
  \BibitemOpen
  \bibfield  {author} {\bibinfo {author} {\bibfnamefont {M.}~\bibnamefont
  {Ivanova}}, \bibinfo {author} {\bibfnamefont {S.}~\bibnamefont {Ricote}},
  \bibinfo {author} {\bibfnamefont {W.~A.}\ \bibnamefont {Meulenberg}},
  \bibinfo {author} {\bibfnamefont {R.}~\bibnamefont {Haugsrud}},\ and\
  \bibinfo {author} {\bibfnamefont {M.}~\bibnamefont {Ziegner}},\ }\bibfield
  {title} {\bibinfo {title} {Effects of a-and b-site (co-) acceptor doping on
  the structure and proton conductivity of lanbo4},\ }\href
  {https://doi.org/10.1016/j.ssi.2011.06.012} {\bibfield  {journal} {\bibinfo
  {journal} {Solid State Ionics}\ }\textbf {\bibinfo {volume} {213}},\ \bibinfo
  {pages} {45} (\bibinfo {year} {2012})}\BibitemShut {NoStop}%
\bibitem [{\citenamefont {Li}\ \emph {et~al.}(2011)\citenamefont {Li},
  \citenamefont {Mori}, \citenamefont {Ye}, \citenamefont {Ou}, \citenamefont
  {Zou},\ and\ \citenamefont {Drennan}}]{li2011structural}%
  \BibitemOpen
  \bibfield  {author} {\bibinfo {author} {\bibfnamefont {Z.-P.}\ \bibnamefont
  {Li}}, \bibinfo {author} {\bibfnamefont {T.}~\bibnamefont {Mori}}, \bibinfo
  {author} {\bibfnamefont {F.}~\bibnamefont {Ye}}, \bibinfo {author}
  {\bibfnamefont {D.~R.}\ \bibnamefont {Ou}}, \bibinfo {author} {\bibfnamefont
  {J.}~\bibnamefont {Zou}},\ and\ \bibinfo {author} {\bibfnamefont
  {J.}~\bibnamefont {Drennan}},\ }\bibfield  {title} {\bibinfo {title}
  {Structural phase transformation through defect cluster growth in gd-doped
  ceria},\ }\href {https://doi.org/10.1103/PhysRevB.84.180201} {\bibfield
  {journal} {\bibinfo  {journal} {Physical Review B—Condensed Matter and
  Materials Physics}\ }\textbf {\bibinfo {volume} {84}},\ \bibinfo {pages}
  {180201} (\bibinfo {year} {2011})}\BibitemShut {NoStop}%
\bibitem [{\citenamefont {Bl{\"o}chl}(1994)}]{blochl1994projector}%
  \BibitemOpen
  \bibfield  {author} {\bibinfo {author} {\bibfnamefont {P.~E.}\ \bibnamefont
  {Bl{\"o}chl}},\ }\bibfield  {title} {\bibinfo {title} {Projector
  augmented-wave method},\ }\href {https://doi.org/10.1103/PhysRevB.50.17953}
  {\bibfield  {journal} {\bibinfo  {journal} {Physical review B}\ }\textbf
  {\bibinfo {volume} {50}},\ \bibinfo {pages} {17953} (\bibinfo {year}
  {1994})}\BibitemShut {NoStop}%
\bibitem [{\citenamefont {Perdew}\ \emph {et~al.}(1996)\citenamefont {Perdew},
  \citenamefont {Burke},\ and\ \citenamefont
  {Ernzerhof}}]{perdew1996generalized}%
  \BibitemOpen
  \bibfield  {author} {\bibinfo {author} {\bibfnamefont {J.~P.}\ \bibnamefont
  {Perdew}}, \bibinfo {author} {\bibfnamefont {K.}~\bibnamefont {Burke}},\ and\
  \bibinfo {author} {\bibfnamefont {M.}~\bibnamefont {Ernzerhof}},\ }\bibfield
  {title} {\bibinfo {title} {Generalized gradient approximation made simple},\
  }\href {https://doi.org/10.1103/PhysRevLett.77.3865} {\bibfield  {journal}
  {\bibinfo  {journal} {Physical review letters}\ }\textbf {\bibinfo {volume}
  {77}},\ \bibinfo {pages} {3865} (\bibinfo {year} {1996})}\BibitemShut
  {NoStop}%
\bibitem [{\citenamefont {Kresse}\ and\ \citenamefont
  {Furthm{\"u}ller}(1996)}]{kresse1996efficient}%
  \BibitemOpen
  \bibfield  {author} {\bibinfo {author} {\bibfnamefont {G.}~\bibnamefont
  {Kresse}}\ and\ \bibinfo {author} {\bibfnamefont {J.}~\bibnamefont
  {Furthm{\"u}ller}},\ }\bibfield  {title} {\bibinfo {title} {Efficient
  iterative schemes for ab initio total-energy calculations using a plane-wave
  basis set},\ }\href {https://doi.org/10.1103/PhysRevB.54.11169} {\bibfield
  {journal} {\bibinfo  {journal} {Physical review B}\ }\textbf {\bibinfo
  {volume} {54}},\ \bibinfo {pages} {11169} (\bibinfo {year}
  {1996})}\BibitemShut {NoStop}%
\bibitem [{\citenamefont {Min}\ \emph {et~al.}(2022)\citenamefont {Min},
  \citenamefont {Yang}, \citenamefont {Zhong},\ and\ \citenamefont
  {Lu}}]{min2022first}%
  \BibitemOpen
  \bibfield  {author} {\bibinfo {author} {\bibfnamefont {Z.}~\bibnamefont
  {Min}}, \bibinfo {author} {\bibfnamefont {C.}~\bibnamefont {Yang}}, \bibinfo
  {author} {\bibfnamefont {G.-H.}\ \bibnamefont {Zhong}},\ and\ \bibinfo
  {author} {\bibfnamefont {Z.}~\bibnamefont {Lu}},\ }\bibfield  {title}
  {\bibinfo {title} {First-principles insights into lithium-rich ternary
  phosphide superionic conductors: Solid electrolytes or active electrodes},\
  }\href {https://doi.org/10.1021/acsami.2c00292} {\bibfield  {journal}
  {\bibinfo  {journal} {ACS Applied Materials \& Interfaces}\ }\textbf
  {\bibinfo {volume} {14}},\ \bibinfo {pages} {18373} (\bibinfo {year}
  {2022})}\BibitemShut {NoStop}%
\bibitem [{\citenamefont {Monkhorst}\ and\ \citenamefont
  {Pack}(1976)}]{monkhorst1976special}%
  \BibitemOpen
  \bibfield  {author} {\bibinfo {author} {\bibfnamefont {H.~J.}\ \bibnamefont
  {Monkhorst}}\ and\ \bibinfo {author} {\bibfnamefont {J.~D.}\ \bibnamefont
  {Pack}},\ }\bibfield  {title} {\bibinfo {title} {Special points for
  brillouin-zone integrations},\ }\href
  {https://doi.org/10.1103/PhysRevB.13.5188} {\bibfield  {journal} {\bibinfo
  {journal} {Physical review B}\ }\textbf {\bibinfo {volume} {13}},\ \bibinfo
  {pages} {5188} (\bibinfo {year} {1976})}\BibitemShut {NoStop}%
\bibitem [{\citenamefont {Yu}\ \emph {et~al.}(2009)\citenamefont {Yu},
  \citenamefont {Lin}, \citenamefont {Wang}, \citenamefont {Chen},\ and\
  \citenamefont {Huang}}]{yu2009first}%
  \BibitemOpen
  \bibfield  {author} {\bibinfo {author} {\bibfnamefont {J.}~\bibnamefont
  {Yu}}, \bibinfo {author} {\bibfnamefont {X.}~\bibnamefont {Lin}}, \bibinfo
  {author} {\bibfnamefont {J.}~\bibnamefont {Wang}}, \bibinfo {author}
  {\bibfnamefont {J.}~\bibnamefont {Chen}},\ and\ \bibinfo {author}
  {\bibfnamefont {W.}~\bibnamefont {Huang}},\ }\bibfield  {title} {\bibinfo
  {title} {First-principles study of the relaxation and energy of bcc-fe,
  fcc-fe and aisi-304 stainless steel surfaces},\ }\href
  {https://doi.org/10.1016/j.apsusc.2009.06.087} {\bibfield  {journal}
  {\bibinfo  {journal} {Applied surface science}\ }\textbf {\bibinfo {volume}
  {255}},\ \bibinfo {pages} {9032} (\bibinfo {year} {2009})}\BibitemShut
  {NoStop}%
\bibitem [{\citenamefont {Henkelman}\ \emph {et~al.}(2000)\citenamefont
  {Henkelman}, \citenamefont {Uberuaga},\ and\ \citenamefont
  {J{\'o}nsson}}]{henkelman2000climbing}%
  \BibitemOpen
  \bibfield  {author} {\bibinfo {author} {\bibfnamefont {G.}~\bibnamefont
  {Henkelman}}, \bibinfo {author} {\bibfnamefont {B.~P.}\ \bibnamefont
  {Uberuaga}},\ and\ \bibinfo {author} {\bibfnamefont {H.}~\bibnamefont
  {J{\'o}nsson}},\ }\bibfield  {title} {\bibinfo {title} {A climbing image
  nudged elastic band method for finding saddle points and minimum energy
  paths},\ }\href {https://doi.org/10.1063/1.1329672} {\bibfield  {journal}
  {\bibinfo  {journal} {The Journal of chemical physics}\ }\textbf {\bibinfo
  {volume} {113}},\ \bibinfo {pages} {9901} (\bibinfo {year}
  {2000})}\BibitemShut {NoStop}%
\bibitem [{\citenamefont {Guevara}\ \emph {et~al.}(1998)\citenamefont
  {Guevara}, \citenamefont {Cuffini}, \citenamefont {Mascarenhas},
  \citenamefont {Carbonio}, \citenamefont {Alonso}, \citenamefont {Fernandez},
  \citenamefont {De~La~Presa}, \citenamefont {Ayala-Morales},\ and\
  \citenamefont {Lopez~Garcia}}]{guevara1998structure}%
  \BibitemOpen
  \bibfield  {author} {\bibinfo {author} {\bibfnamefont {J.}~\bibnamefont
  {Guevara}}, \bibinfo {author} {\bibfnamefont {S.}~\bibnamefont {Cuffini}},
  \bibinfo {author} {\bibfnamefont {Y.~P.}\ \bibnamefont {Mascarenhas}},
  \bibinfo {author} {\bibfnamefont {R.}~\bibnamefont {Carbonio}}, \bibinfo
  {author} {\bibfnamefont {J.~A.}\ \bibnamefont {Alonso}}, \bibinfo {author}
  {\bibfnamefont {M.}~\bibnamefont {Fernandez}}, \bibinfo {author}
  {\bibfnamefont {P.}~\bibnamefont {De~La~Presa}}, \bibinfo {author}
  {\bibfnamefont {A.}~\bibnamefont {Ayala-Morales}},\ and\ \bibinfo {author}
  {\bibfnamefont {A.}~\bibnamefont {Lopez~Garcia}},\ }\bibfield  {title}
  {\bibinfo {title} {The structure of orthorhombic hafniates by neutron powder
  diffraction and perturbed-angular-correlation spectroscopy (pac)},\ }in\
  \href@noop {} {\emph {\bibinfo {booktitle} {Materials Science Forum}}},\
  Vol.\ \bibinfo {volume} {278}\ (\bibinfo {organization} {Trans Tech Publ},\
  \bibinfo {year} {1998})\ pp.\ \bibinfo {pages} {720--725}\BibitemShut
  {NoStop}%
\bibitem [{\citenamefont {Lee}\ \emph {et~al.}(2022)\citenamefont {Lee},
  \citenamefont {Duan}, \citenamefont {Sorescu}, \citenamefont {Saidi},
  \citenamefont {Morgan}, \citenamefont {Thomas}, \citenamefont {Epting},
  \citenamefont {Hackett},\ and\ \citenamefont {Abernathy}}]{lee2022defect}%
  \BibitemOpen
  \bibfield  {author} {\bibinfo {author} {\bibfnamefont {Y.-L.}\ \bibnamefont
  {Lee}}, \bibinfo {author} {\bibfnamefont {Y.}~\bibnamefont {Duan}}, \bibinfo
  {author} {\bibfnamefont {D.~C.}\ \bibnamefont {Sorescu}}, \bibinfo {author}
  {\bibfnamefont {W.~A.}\ \bibnamefont {Saidi}}, \bibinfo {author}
  {\bibfnamefont {D.}~\bibnamefont {Morgan}}, \bibinfo {author} {\bibfnamefont
  {K.}~\bibnamefont {Thomas}}, \bibinfo {author} {\bibfnamefont {W.~K.}\
  \bibnamefont {Epting}}, \bibinfo {author} {\bibfnamefont {G.}~\bibnamefont
  {Hackett}},\ and\ \bibinfo {author} {\bibfnamefont {H.}~\bibnamefont
  {Abernathy}},\ }\bibfield  {title} {\bibinfo {title} {Defect thermodynamics
  and transport properties of proton conducting oxide bazr1-xyxo3- $\delta$
  (x$\leq$ 0.1) guided by density functional theory modeling},\ }\href
  {https://doi.org/10.1007/s11837-022-05554-z} {\bibfield  {journal} {\bibinfo
  {journal} {JOM}\ }\textbf {\bibinfo {volume} {74}},\ \bibinfo {pages} {4506}
  (\bibinfo {year} {2022})}\BibitemShut {NoStop}%
\bibitem [{\citenamefont {Ganduglia-Pirovano}\ and\ \citenamefont
  {Sauer}(2004)}]{ganduglia2004stability}%
  \BibitemOpen
  \bibfield  {author} {\bibinfo {author} {\bibfnamefont {M.}~\bibnamefont
  {Ganduglia-Pirovano}}\ and\ \bibinfo {author} {\bibfnamefont
  {J.}~\bibnamefont {Sauer}},\ }\bibfield  {title} {\bibinfo {title} {Stability
  of reduced v 2 o 5 (001) surfaces},\ }\href
  {https://doi.org/10.1103/PhysRevB.70.045422} {\bibfield  {journal} {\bibinfo
   {journal} {Physical Review B—Condensed Matter and Materials Physics}\
  }\textbf {\bibinfo {volume} {70}},\ \bibinfo {pages} {045422} (\bibinfo
  {year} {2004})}\BibitemShut {NoStop}%
\bibitem [{\citenamefont {Feng}\ \emph {et~al.}(2025)\citenamefont {Feng},
  \citenamefont {Ma}, \citenamefont {Yang}, \citenamefont {Lv}, \citenamefont
  {Liang}, \citenamefont {Ma}, \citenamefont {Linghu},\ and\ \citenamefont
  {Li}}]{Feng2025}%
  \BibitemOpen
  \bibfield  {author} {\bibinfo {author} {\bibfnamefont {P.}~\bibnamefont
  {Feng}}, \bibinfo {author} {\bibfnamefont {H.}~\bibnamefont {Ma}}, \bibinfo
  {author} {\bibfnamefont {K.}~\bibnamefont {Yang}}, \bibinfo {author}
  {\bibfnamefont {Y.}~\bibnamefont {Lv}}, \bibinfo {author} {\bibfnamefont
  {Y.}~\bibnamefont {Liang}}, \bibinfo {author} {\bibfnamefont
  {T.}~\bibnamefont {Ma}}, \bibinfo {author} {\bibfnamefont {J.}~\bibnamefont
  {Linghu}},\ and\ \bibinfo {author} {\bibfnamefont {Z.-P.}\ \bibnamefont
  {Li}},\ }\bibfield  {title} {\bibinfo {title} {Data associated with in-depth
  investigation of conduction mechanism on defect-induced proton-conducting
  electrolytes bahfo3},\ }\href {https://doi.org/10.5281/zenodo.15422695}
  {10.5281/zenodo.15422695} (\bibinfo {year} {2025})\BibitemShut {NoStop}%
\end{thebibliography}%

\end{document}